\PassOptionsToPackage{table}{xcolor}
\documentclass{article}

     \usepackage[preprint,nonatbib]{neurips_2025}

\usepackage[utf8]{inputenc} % allow utf-8 input
\usepackage[T1]{fontenc}    % use 8-bit T1 fonts
\usepackage{hyperref}       % hyperlinks
\usepackage{url}            % simple URL typesetting
\usepackage{booktabs}       % professional-quality tables
\usepackage{amsfonts}       % blackboard math symbols
\usepackage{nicefrac}       % compact symbols for 1/2, etc.
\usepackage{microtype}      % microtypography
\usepackage{xcolor}         % colors
\usepackage{graphicx}
\usepackage{amsmath} 
\usepackage{comment}
 \usepackage{enumitem} % 放在导言区

\title{$\text{M}^3\text{PDB}$: A Multimodal, Multi-Label, Multilingual Prompt Database for Speech Generation}

\author{%
  Boyu Zhu\textsuperscript{1} \And
  Cheng Gong\textsuperscript{1} \And
  Muyang Wu\textsuperscript{2} \And
  Ruihao Jing\textsuperscript{1} \And
  Fan Liu\textsuperscript{1} \And 
  Xiaolei Zhang\textsuperscript{1} \And
  Chi Zhang\textsuperscript{1} \And
  Xuelong Li\textsuperscript{1}\And \\  % <- 作者列表结束后再换行
  \textsuperscript{1}Institute of Artificial Intelligence (TeleAl), China Telecom\\
  \textsuperscript{2}School of Marine Science and Technology, Northwestern Polytechnical University\\
   {\texttt{zhuboyu@mail.nwpu.edu.cn}}
}

\begin{document}

\maketitle

\begin{abstract}
Recent advancements in zero-shot speech generation have enabled models to synthesize speech that mimics speaker identity and speaking style from speech prompts. 
However, these models' effectiveness is significantly limited in real-world scenarios where high-quality speech prompts are absent, incomplete, or out of domain. 
This issue arises primarily from a significant quality mismatch between the speech data utilized for model training and the input prompt speech during inference.
%These challenges are exacerbated when prompts are noisy, entirely missing, or originate from unseen languages. 
To address this, we introduce $\text{M}^3\text{PDB}$, the first large-scale, multi-modal, multi-label, and multilingual prompt database designed for robust prompt selection in speech generation. 
Our dataset construction leverages a novel multi-modal, multi-agent annotation framework, enabling precise and hierarchical labeling across diverse modalities. Furthermore, we propose a lightweight yet effective prompt selection strategy tailored for real-time, resource-constrained inference settings. 
Experimental results demonstrate that our proposed database and selection strategy effectively support various challenging speech generation scenarios. 
We hope our work can inspire the community to shift focus from improving performance on standard benchmarks to addressing more realistic and diverse application scenarios in speech generation.
Code and dataset are available at: \url{https://github.com/hizening/M3PDB}.
\end{abstract}

\section{Introduction}
Recent years have seen significant advancements in zero-shot speech generation research \cite{9054535,10.5555/3692070.3692979,NEURIPS2023_3eaad2a0}, with various models leveraging large-scale training datasets \cite{zen19_interspeech,chen21o_interspeech,pratap20_interspeech}. These advancements showcase the ability to synthesize speech for any speaker by mimicking the timbre, prosody, and style of a given speech prompt. 
However, in real-world scenarios, high-quality reference prompts are often unavailable, which significantly impacts the quality of the generated speech. When real-time input is insufficient or inappropriate as a prompt, it is necessary to select an appropriate audio sample from the existing database as a substitute.

This issue mainly stems from a significant quality mismatch between the speech data used for model training and the input prompt speech during inference.
Despite recent progress in large-scale and high-quality speech datasets \cite{zen19_interspeech,chen21o_interspeech,pratap20_interspeech}, existing resources are predominantly collected under ideal conditions, such as professional studios or curated audiobook recordings. While these datasets benefit training models, they often fail to handle complex inference scenarios.
Specifically, there are three common scenarios where prompt selection becomes particularly challenging:
\begin{itemize}[leftmargin=*]
    \item \textbf{Low-quality speech prompts:} User-provided audio recordings are often captured in noisy, unconstrained environments, leading to degraded prompt quality. Moreover, streaming speech synthesis systems frequently operate on partial or incomplete input windows, further complicating the use of speech prompts.
    \item \textbf{Absence of speech prompts:} In many real-world settings, users may only provide visual inputs (e.g., facial images) or textual descriptions of the desired voice characteristics, without any accompanying speech data. 
    \item \textbf{Unseen languages: }A growing number of multilingual applications require synthesis in languages that are underrepresented or absent from the training data. When such unseen languages are used directly as prompts, the synthesis quality often deteriorates due to the model’s limited generalization capability.
\end{itemize}

These challenges highlight the urgent need for a prompt database and selection method that can effectively manage cross-modal inputs, adapt to unseen conditions, and ensure high synthesis fidelity across various inference scenarios. In this study, we create a large-scale, multi-modal, multi-label, and multilingual dataset called $\text{M}^3\text{PDB}$ to address the challenges mentioned above. Additionally, we propose an effective approach for selecting reference prompts with limited computational resources. The key contributions of this paper can be summarized as follows:
\begin{itemize}[leftmargin=*]
    \item To the best of our knowledge, we present the first large-scale dataset specifically designed for prompt selection in speech generation. While constructing this dataset, we introduce a novel multi-modal multi-agent annotation framework, enabling accurate and hierarchical annotations across various data modalities.
    \item Our database enables robust prompt selection under challenging conditions, such as low-quality input audio or visual-only inputs.
    \item  Combined with our dataset, we propose an efficient prompt selection strategy that enables fast and resource-efficient prompt selection under real-time and resource-constrained inference conditions.
\end{itemize}

\section{Related Works}
\label{gen_inst}
\paragraph{Speech Generation Datasets}
Over the years, numerous corpora for speech generation have been released. 
Table \ref{tab:speech_datasets} compares our $\text{M}^3\text{PDB}$ datasets with several existing speech generation datasets.
Current speech datasets mainly focus on scaling up datasets and enhancing the granularity of annotations.
The size of speech datasets has increased from tens to hundreds of hours, and more recently, to thousands or even tens of thousands of hours.
For instance, LibriTTS \cite{zen19_interspeech} comprises 585.8 hours of audiobook speech, whereas GigaSpeech \cite{chen21o_interspeech} provides 10,000 hours of raw, in-the-wild speech data from a wide range of sources.
Moreover, speech datasets are increasingly multilingual; for instance, the MLS \cite{pratap20_interspeech} dataset includes eight commonly spoken Indo-European languages.

Early datasets typically contained only basic annotations such as speaker identity and gender. Recently, SpeechCraft \cite{10.1145/3664647.3681674} has been proposed, a fine-grained expressive speech dataset with eight labels. However, both the scale and annotation accuracy of SpeechCraft \cite{10.1145/3664647.3681674} remain to be improved. More importantly, all the above speech datasets include only two modalities—speech and text, which significantly constrain the generative capabilities of models. 
The recent advancement of large models \cite{chu2023qwen,an2024funaudiollm,cheng2024videollama} for speech and video understanding has enabled the automatic annotation of audio-visual data from various perspectives. 
However, choosing the most appropriate model for a particular annotation task continues to be a significant challenge.  

\begin{table*}[htbp]
    \centering
    \caption{A comparison of $\text{M}^3\text{PDB}$ with existing datasets for speech generation. T, A, and V represent the three modalities: Text, Audio, and Visual.}
    \label{tab:speech_datasets}
    \resizebox{\textwidth}{!}{
        \begin{tabular}{lcccccc}
            \toprule
            \textbf{Dataset} & \textbf{Total Duration (hours)} & \textbf{Languages} & \textbf{Modality} & \textbf{Speakers}  & \textbf{Label Types} \\
            \midrule
            \textbf{LibriTTS~\cite{zen19_interspeech}} & 586 & 1 & T+A & 2,456 & 3  \\
            \textbf{GigaSpeech~\cite{chen21o_interspeech}} & 10,000 & 1 & T+A & N/A &  5 \\
            \textbf{WenetSpeech4TTS~\cite{ma24d_interspeech}} & 12,800  & 1 & T+A & N/A & 3\\
            \textbf{MLS~\cite{pratap20_interspeech}} & 50,500 & 8 & T+A & 6,332 &  3\\ 
            \textbf{SpeechCraft~\cite{10.1145/3664647.3681674}} & 2,931 & 2  &T+A & >3,200  &  8\\
            \textbf{Emila~\cite{he2025emilia}} & 101,654 & 6 & T+A  & N/A  & 6 \\
            \midrule
            \rowcolor{gray!20} 
            \textbf{$\text{M}^3\text{PDB}$ (Our)} & \textbf{403,308} &  \textbf{18} & \textbf{T+A+V} &  \textbf{>15,000}  & \textbf{10} \\
            \bottomrule
        \end{tabular}
    }
    \vspace{-0.5cm}
\end{table*}

%an open-source, large-scale bilingual dataset available for advanced speech-language learning with fine-grained and expressive descriptions of speech,
%It is widely acknowledged that human annotation datasets are typically costly, time-consuming, and limited in scope.

\paragraph{Speech Generation Model}
Using large amounts of high-quality labeled data during training is essential for significant advancements in the speech generation task, such as zero-shot TTS (Text-to-speech) and S2ST (Speech-to-speech translation). 
Zero-shot TTS \cite{9054535,10.5555/3692070.3692979,NEURIPS2023_3eaad2a0} aims to synthesize unseen voices with speech prompts.  
Recent advancements in large language models (LLMs) have driven significant progress in zero-shot TTS  synthesis.
These LLM-based works \cite{10842513,du2024cosyvoice,anastassiou2024seed,wang2025spark,chen2024vall} typically convert text input into quantized representations derived from a pretrained neural speech codec. 
Among them, neural codec language models \cite{chen2024vall}  are the first to autoregressively synthesize speech that matches human recordings regarding naturalness and expressiveness. 
Inspired by the natural ability of people to imagine someone's voice when they see their face, some studies \cite{10094745,lee24_interspeech} have explored the use of visual cues, such as facial images, as prompts for zero-shot TTS instead of using audio. While image-based references provide additional flexibility in selecting prompts, the quality of the synthesized speech tends to be low due to the limited amount of paired audio-visual data available.

As zero-shot TTS technology rapidly advances, more studies are focusing on other speech generation tasks, such as S2ST \cite{barrault2023seamless,NEURIPS2024_a32539cb} and speech interaction \cite{fu2025vita}, to preserve speaker timbre and vocal style. For example, TransVIP \cite{NEURIPS2024_a32539cb} proposes two separate encoders to preserve the speaker’s voice characteristics from the source speech during the translation process. 
However, inference scenarios in real-world applications can be complex and often face challenges, such as missing or poor-quality prompt speech. There is an urgent need to develop a comprehensive prompt speech database to utilize the model's generative potential fully.

%In this paper, we develop a speech-to-speech translation framework that preserves speaker voice and isochrony, generating high-quality translated speech suitable for both daily communication and automatic video dubbing.
%Recent advancements in large language models (LLMs) have driven significant progress in zero-shot text-to-speech (TTS) synthesis. Recent advances in speech tokenization have revolutionized text-to-speech (TTS) synthesis by bridging the fundamental gap between continuous speech signals and discrete token-based large language models (LLMs) (Anastassiou et al., 2024; Zhu et al., 2024; Wang et al., 2024c).
%There is an urgent need for
%Using large amounts of high-quality labeled data during training is essential for significant advancements in the speech generation task, such as zero-shot TTS.

\paragraph{Prompt Selection Strategy} 
As previously mentioned, selecting an appropriate speech reference is crucial for prompt-based speech generation \cite{wang2024emopro}. However, there has been limited previous research focusing on the importance of prompt selection.
Currently, most approaches \cite{9747801,luo2025autostyle} select suitable reference audio by matching the text's semantic content with the audio's style.
For instance, the study \cite{luo2025autostyle} proposes
a TTS framework based on Retrieval-Augmented
Generation (RAG) technology, which can dynamically adjust the speech style according to the text content to achieve more
natural and vivid communication effects.
On the one hand, constrained by the limited size of reference audio repositories, current selection strategies have made little breakthrough. On the other hand, relying solely on text for selection struggles to handle \textbf{complex inference scenarios}, such as those involving streaming or emotional synthesis. Notably, by leveraging our large-scale prompt speech database and fast retrieval strategy, we enable speech synthesis models to handle a wide range of complex inference scenarios.

\subsection{Overview of Data Collection and Annotation}
Considering a comprehensive set of factors such as dataset audio and image quality, the number and distribution of speakers and languages, we construct speech generation datasets derived from a set of public data and synthesized data. 
To enhance data quality, we first perform data preprocessing. 
Given processed data, we proposed a novel automatic speech annotation system for prompt selection in speech generation. 
%For speakers with only a few audio samples, we further augment their data by synthesizing additional speech using a zero-shot TTS model.
For speakers with limited audio samples, we agument their data by generating additional speech using a zero-shot TTS model CosyVoice\footnote{https://github.com/FunAudioLLM/CosyVoice}.
\textbf{The $\text{M}^3\text{PDB}$ dataset includes annotations across 10 labels for both speech and visual modalities, covering 18 languages, with about 15k speakers and a total of 400k hours of data.}
Comprehensive information about the data distribution and preprocessing methods can be found in \textcolor{black}{Appendix A.}

\begin{figure}[]
  \centering
  %\vspace{-0.65cm}
  \includegraphics[width=0.8\textwidth]{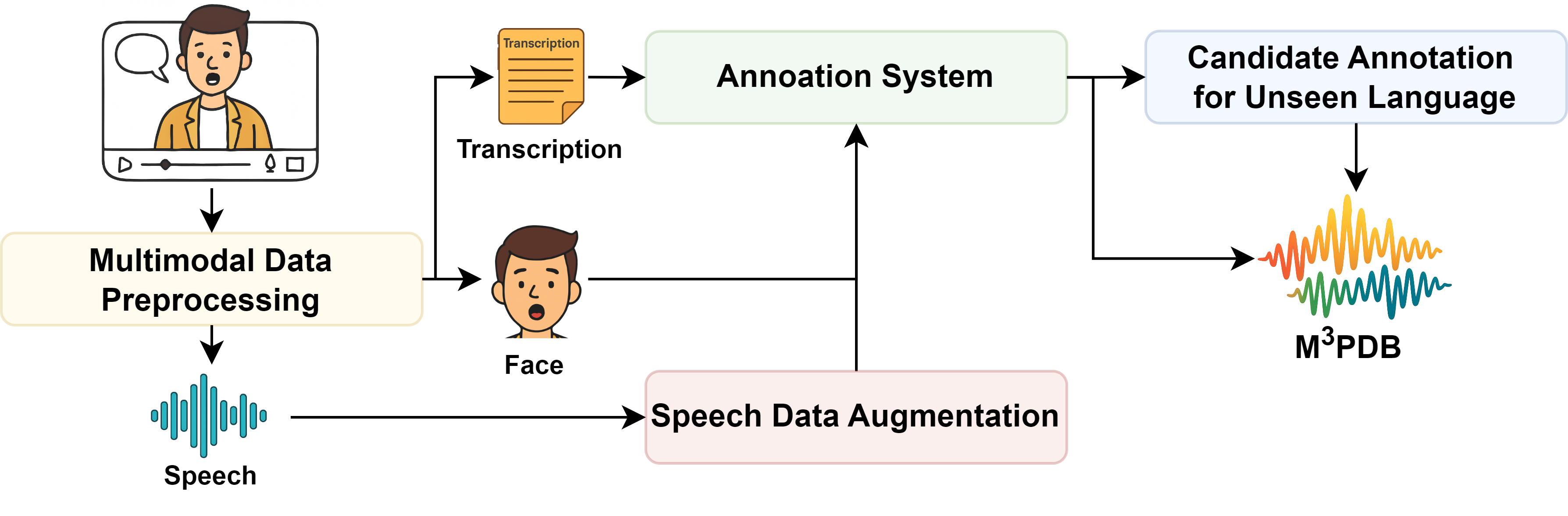}
  %\vspace{-0.9cm}
  \caption{The overall framework for data collection and annotation.}
  \label{fig:anno}
  %\vspace{-0.5cm}
\end{figure}

The overall framework for the data collection and annotation is illustrated in Figure \ref{fig:anno}. 
In this section, we present a detailed introduction to our multi-modal and multi-agent annotation approach, along with our annotation strategy for unseen languages, which is key to addressing unseen language in speech generation.
%In section 4, we will discuss the codec and NAR model.

\begin{comment}
Raw recordings of audio and audio–video streams from diverse sources differ markedly in both quality and original metadata formats. To standardize these data into an enriched format, we first perform preprocessing to boost overall quality. Specifically, for audio signals, we adopt the preprocessing pipeline supplied by Emilia.
For video streams, we begin by using FFmpeg to demultiplex the audio and video tracks. The extracted audio then undergoes the same preprocessing steps as before, while the video is processed through cropping and temporal alignment, followed by quality restoration and frontal‐face detection. This workflow yields high‐resolution, frontal facial images accurately synchronized with the corresponding speech.
\end{comment}
\section{$\text{M}^3\text{PDB}$ Dataset}

\subsection{Multi-modal Multi-agent Annotation}
\begin{figure}[]
  \centering
  %\vspace{-0.65cm}
  \includegraphics[width=\textwidth]{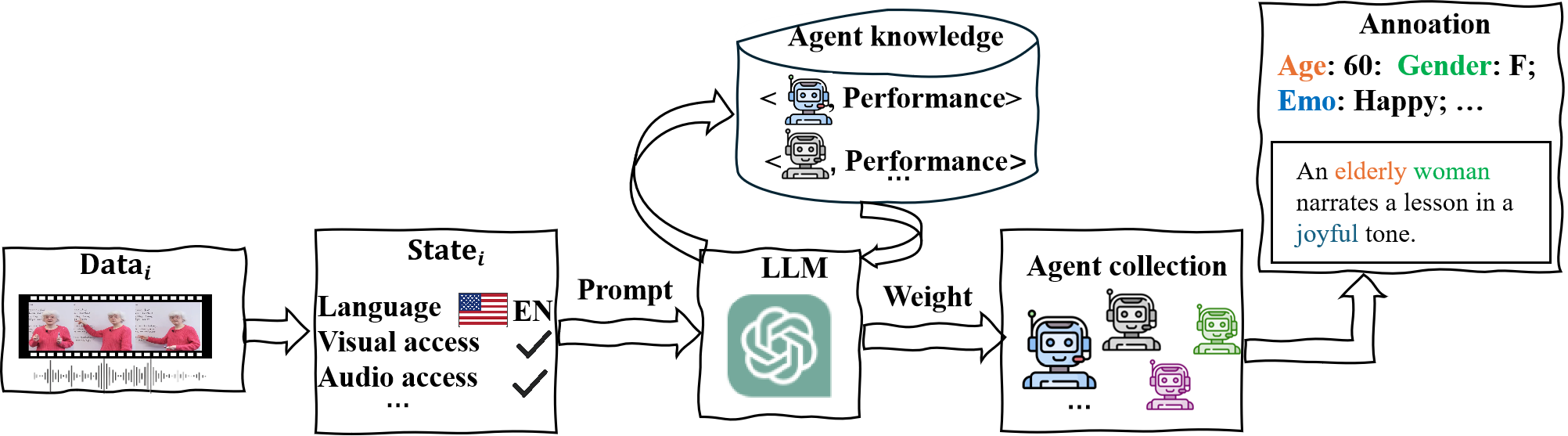}
  %\vspace{-0.9cm}
  \caption{The structure of multi-modal multi-agent annotation.}
  \label{fig:}
  \label{fig:2}
  %\vspace{-0.5cm}
\end{figure}
The characteristics of a multi-modal multi-agent system include: 1) an LLM-based central controller with the RAG as the central control unit for individual agents; 2) a collection of agent models across different modalities; and 3) a hierarchical annotation module.

To allow the LLM to automatically select the correct agent model for a specific annotation task, it is essential to first build a knowledge base $K$ of available agent models. This knowledge base encodes the recognition performance of different agent models across various tasks and supported languages.
Each data instance, \( Data_i \), that needs to be annotated is first transformed into a structured state description, \( Satae_i \). This description includes attributes such as the language and the presence of visual content. Specifically, we incorporate the following instruction into the prompt: \textit{"Please select appropriate annotation models and assign weights based on their performance in the knowledge base $K$ for the given data instance (\( Satae_i \))."} The LLM then queries the knowledge base $K$ using the state description $Satae_i$, and outputs a weight $AW_j$ for each relevant agent model $A_j$. 
Finally, the annotation result for each task is obtained by aggregating the outputs of different agent models in a weighted manner.
Figure \ref{fig:2} illustrates a sample annotation process using our multimodal and multi-agent framework. The detailed annotation process, as well as the configurations of the LLM and various agent models, can be found in \textcolor{black}{Appendix B.}

In addition to using traditional discrete labels, we also utilize hierarchical annotations. First, inspired by the SpeechCraft approach, we employ a large language model (LLM) to transform discrete labels into natural language descriptions. Furthermore, we pre-extract high-dimensional acoustic representations, such as emotion and speaker embeddings, to serve as auxiliary annotations, which help speed up the prompt selection process.

\begin{comment}
To accelerate and improve the speech‐data labeling task, we propose a two‐stage system: first, a RAG‐augmented LLM serves as a sparse gating network that dynamically selects among modality‐specific experts; second, a general‐purpose LLM combines their outputs into fluent text. The overall architecture is shown in Figure 2.

Selecting the best expert for each utterance is nontrivial.  For example, in zero‐shot emotion tagging on an unseen language, a dedicated emotion classifier often underperforms a simple pitch‐based heuristic.  Conversely, for age and gender prediction, vision models consistently outstrip audio‐only networks.  To guide the gating decision, we augment the RAG‐LLM with each expert’s benchmark performance as an external knowledge source, enabling it to route inputs to the most reliable model.

Our expert pool spans three modalities—speech understanding, text understanding, and image understanding—so that each labeler can leverage its strongest signal.  Finally, a unified LLM ingests all expert annotations and generates a coherent, human‐readable description.
\end{comment}

\subsection{Candidate Annotation for Unseen Language}
\label{sec:3_3}
To extend our $\text{M}^3\text{PDB}$ to support speech generation in more languages, we annotate a subset of samples in the $\text{M}^3\text{PDB}$ dataset as candidate prompts for those languages not included in the $\text{M}^3\text{PDB}$.
%We design a set of criteria, and only the audio samples that meet these criteria are labeled as candidate prompts for the corresponding unseen language. 
First, at the language level, we select the closest language $B$ as a proxy language for an unseen language $A$ based on the distance in the language family tree.
All speech samples from language B are further evaluated based on two aspects: accent and quality. 
For accent evaluation, we propose a metric for accent similarity based on probabilities from language identification (LID) with a speech recognition model whisper\footnote{https://github.com/openai/whisper}.
%Using a zero-shot speech generation model $\text{TTS}()$,  we can generate speech $A_{Speech}$ in language $A$ follow a cross-lingual synthesis as $A_{Speech} = \text{TTS}(A_{Text}, B_{Speech})$. $A_{Text}$ is the target synthesized text in language, while $B_{Speech}$ is the reference prompt in language $B$.
By utilizing a zero-shot speech generation model, denoted as $\text{TTS}()$, we can generate speech denoted as $A_{Speech}$ in language $A$ through a process of cross-lingual synthesis. 
This process can be expressed as: $A_{Speech} = \text{TTS}(A_{Text}, B_{Speech})$, where $A_{Text}$ represents the target synthesized text in language $A$, while $B_{Speech}$ serves as the reference prompt in language $B$ from $\text{M}^3\text{PDB}$.
The $A_{Speech}$ is processed through a LID model, and only those samples where the predicted probability of language A exceeds 95\% are retained as potential prompts. Finally, we conduct objective evaluations of the $A_{Speech}$ from multiple angles, including similarity and content quality. Only the samples that meet all predefined threshold criteria are annotated as valid candidates. The detailed criteria computation process can be referred to in \textcolor{black}{Appendix C}. 
\begin{figure}[]
  \centering
  %\vspace{-0.65cm}
  \includegraphics[width=0.73\textwidth]{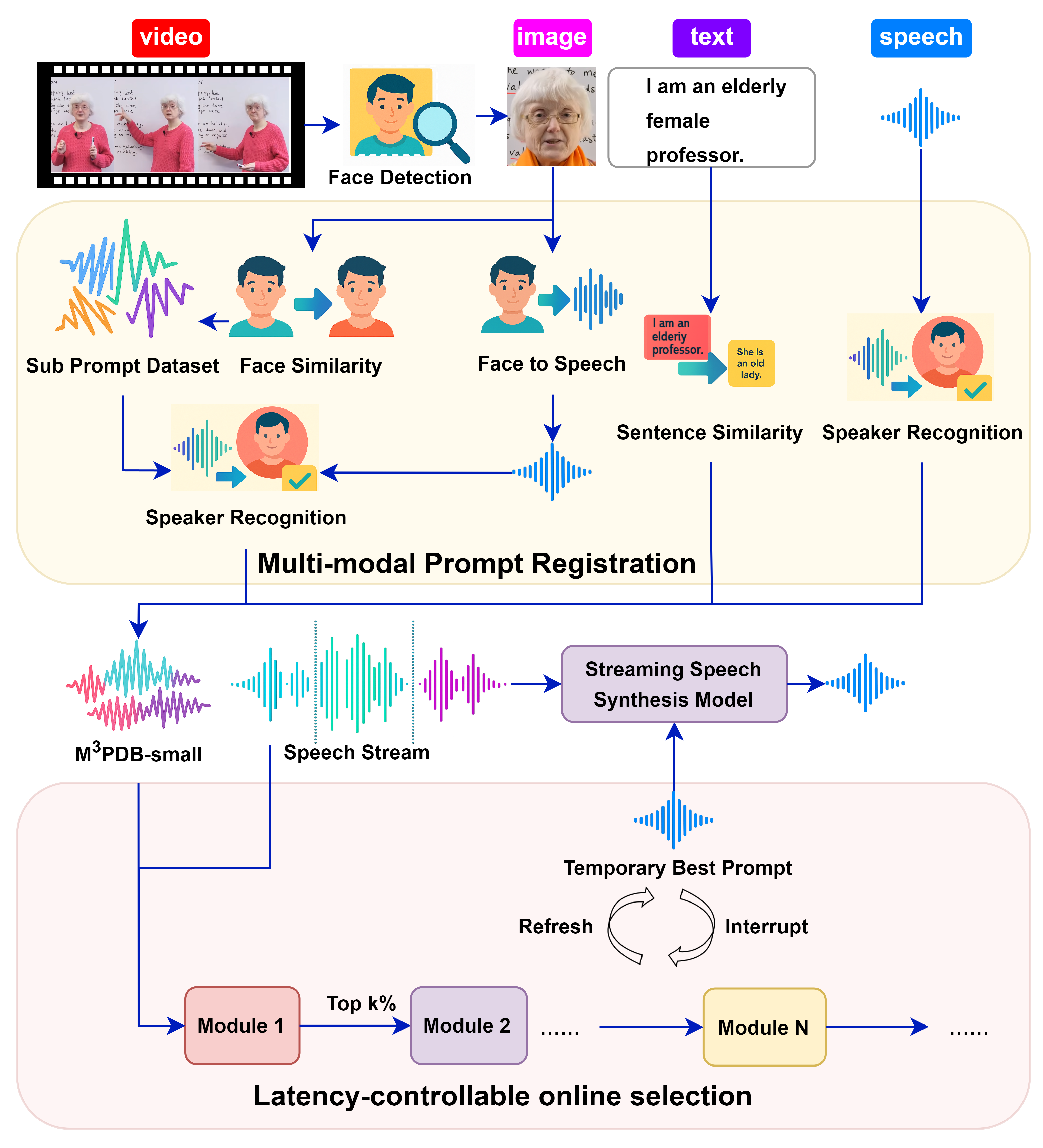}
  %\vspace{-0.9cm}
  \caption{The overall architecture of latency-aware prompt selection.}
  \label{fig:3}
  \vspace{-0.5cm}
\end{figure}
\section{Latency-aware Prompt Selection}
To address the varying latency needs of different applications, we propose a two-stage prompt selection strategy. This strategy consists of an offline multi-model prompt registration and a real-time (or online) selection phase. In the offline registration stage, modality-specific registration information is used to narrow down the reference prompt search space by selecting a smaller candidate subset $\text{M}^3\text{PDB}_{small}$ for use in subsequent real-time selection. The overall architecture of Latency-aware prompt selection is shown in Figure \ref{fig:3}.
\subsection{Multi-model Prompt Registration}
\label{sec:4_1}
%在很多情况下，获得说话人的真实语音是很困难的，比如游戏或动漫中的某个人物本来就不存在对应的真实语音。为了解决该问题，我们设计了一种跨模态的参考语音选取策略，具体结构如图3所示。首先，由用户给出他对说话人特征的文本描述，或者说话人的视频，或者说话人的图片。我们根据先前得到的标注，在数据库中匹配到一批候选音频。然后通过描述生语音模型或者人脸生语音模型来得到一个引导语音。最后，根据该引导语音在数据库中进行声纹匹配，从而得到的最终的参考说话人音频。
%To address situations where reference speech is absent or of low quality, we develop a strategy that selects suitable reference audio from the dataset using textual or visual information. 
The user can provide a description of the desired voice timbre $T_{reg}$, an audio sample of their voice $A_{reg}$, or a personal image $V_{reg}$. 
Since $\text{M}^3\text{PDB}$ includes natural language descriptions for speech samples, users can simply describe the desired voice characteristics $T_{reg}$—such as\textit{ 'A young woman, voice high, pace swift, revealed joy and delight in her emotion'}—and retrieve the semantically similar reference speech from the database as candidate subset $\text{M}^3\text{PDB}_{small}$.

Given that directly using facial images as prompts often results in low-quality synthesized speech, we propose a two-stage selection strategy for visual input. First, we extract a facial image $F$ from the provided video or image and compute its similarity with facial embeddings in our $\text{M}^3\text{PDB}$. We select faces with the top 20 similarity scores, and their corresponding audio samples form a candidate prompt set. 
In the second stage, we utilize a zero-shot face-to-speech synthesis \textcolor{black}{model}\footnote{https://github.com/naver-ai/facetts} to generate a speech sample $S_f$ that reflects the vocal characteristics inferred from the facial image $F$. We then calculate the similarity between the synthesized sample $S_f$ and each audio sample in the candidate set, choosing the similar as the final candidate subset $\text{M}^3\text{PDB}_{small}$. 
For the speech registration scenario, the selection of candidate set $\text{M}^3\text{PDB}_{small}$ is straightforward: we compute the speaker similarity between each audio sample in the database and the registered speaker A, and then select the most similar samples to form set $\text{M}^3\text{PDB}_{small}$.
For the speech registration scenario, the selection of candidate set $\text{M}^3\text{PDB}_{small}$ is straightforward: we compute the speaker similarity between each audio sample in the database and the registered $A_{reg}$, and then select the similar samples to form $\text{M}^3\text{PDB}_{small}$.
%\textcolor{black}{The detailed process of the multi-model prompt selection can be referred to in Appendix D.}
%\paragraph{Based on text prompt.} 
%实验放demo
%\paragraph{Based on visual prompt.}

\subsection{Latency-aware Online Selection}
%对于不同的语音生成场景而言，对prompt selection的选取策略的时延的要求是多种多样的。对于语音翻译等即时语音交互场景，就需要在1-2s内完成参考音频的选取，并且推理时间限制是不定的，在需要这条参考音频时就得马上。
%\paragraph{Online Selection}
For speech generation tasks with strict time constraints, such as streaming speech translation or real-time speech interaction, prompt selection must be performed efficiently to maintain the synthesis speed of the original pipeline. 
In addition to the time required for similarity computation, it is crucial to consider the limited computational resources. The online selection process should introduce minimal overhead and, ideally, be executable on the same GPU as the speech generation model.

During the multi-model prompt registration phase, the candidate subset \( \text{M}^3\text{PDB}_{small} \) is primarily selected based on time-invariant vocal attributes such as age and gender. In the online selection stage, the final prompt \( P_{final} \) is chosen by assessing the similarity of time-varying characteristics—such as pitch, speaking rate, and emotional tone—between the candidate subset $\text{M}^3\text{PDB}_{small}$ and the real-time input. 

Let $S_i$ denote the $i$-th time-varying characteristics similarity module, where $i \in \{1, 2, \dots, N\}$. Each module computes a specific similarity. Let $T_i$ denote the computational speed of module $S_i$. The final selection time $T_{total}$ could be: 
\begin{equation}
    T_{total} =\sum_{i=1}^{N} T_i*Top_i,  \quad \text{where} \quad Top_1 = 100\%. 
\end{equation}
where $Top_{i}$ is a preset number defined as that, in the similarity computation module $S_i$, similarity is calculated using only the $Top_i$ samples from the $S_{(i-1)}$ similarity computation result. 
In practical applications, the execution order of different similarity computation modules $S_i$ and the size of each $Top_i$ can be flexibly adjusted to meet varying latency requirements. Moreover, the process can be interrupted at any point to return the currently most similar prompt.

\section{Experiments and Results}
In this section, we conduct experiments from the following four perspectives: 1) Verify the advantages of our multi-modal multi-agent annotation strategy in improving data labeling accuracy; 2) Evaluating the validity of the candidate prompt annotation mechanism designed for unseen languages; 3) Assessing the effectiveness of video-based prompt selection; 4) Analyzing the efficiency and performance of the proposed online prompt selection.

\subsection{Multi-modal Multi-agent Annotation}
\label{sec:dataset1}
%为了验证所提出的标注系统的准确性，我们以Voxceleb2，VGGface2为基础，构建了一个用于验证标注准确性的验证数据集。数据集包括了6900条样本对。每个样本对包括了说话人的音频，对应的视频，说话人正脸图片，以及性别、年龄和年龄段的真实标签。
\paragraph{Datasets} To validate the accuracy of our proposed annotation system, we constructed a validation dataset drawn from VoxCeleb2 \cite{chung2018voxceleb2} and VGGFace2 \cite{8373813}. The set comprises 6,900 paired samples, each containing a speaker’s audio, the corresponding video, a frontal face image, and ground-truth labels for gender, age, and age group.
%我们选取了3种客观指标来对标注系统的准确性进行评估。具体如下：
%• 年龄分类准确率（age-acc）：用来评估性别分类的准确性。
%• (age-MAE)：
%. 性别分类准确率(gender-acc):用来评估性别标注的准确性。

\paragraph{Experiment Implementation}
%为了验证我们方法在标注精度方面的优势，我们对比了基于多个专家模型的标注方法Speech Craft，以及一些语音理解大模型，包括了Sense Voice，OSUM等模型。
We compare our proposed multimodal multi-agent annotation method with the SpeechCraft \cite{10.1145/3664647.3681674} approach on the same set of test data, focusing on the accuracy of age and gender annotations. In our approach, an LLM \footnote{https://openai.com/index/gpt-4o-system-card/} assigns output weights across multiple agent models, such as visual\footnote{https://github.com/serengil/deepface}\footnote{https://github.com/nawafalageel/Side-Profile-Detection}, speech \cite{ma2023emotion2vec}, and text \cite{pmlr-v202-longpre23a} modality.
In addition, we compare our method with recent state-of-the-art speech foundation models, such as SenseVoice\footnote{https://github.com/FunAudioLLM/SenseVoice} and OSUM\footnote{https://github.com/ASLP-lab/OSUM}, across a broader set of annotation labels. 
%Detailed results and implementation specifics are in Appendix C.

\paragraph{Metrics}
We adopt the following metrics to evaluate the accuracy of age and gender annotations:
{\textbf{Age-acc}: quantifies the fraction of samples correctly assigned to predefined age groups—Child (<14), Teenager (14–25), Young Adult (26–39), Middle-aged (40–54) and Elderly (>55)—offering a clear measure of age-group discrimination.
\textbf{Age-MAE}: reports the average absolute difference, in years, between predicted and true ages.
\textbf{Gender-acc}: assesses the correctness of gender labels.}
\paragraph{Results}
Table \ref{tab:age_gender_metrics} compares the annotation performance of our proposed method with that of SpeechCraft on the same evaluation set. The results demonstrate the superiority of our multi-modal, multi-agent annotation strategy. While SpeechCraft reports high accuracy for the gender and age classification on its dataset, its performance degrades significantly when applied to more diverse datasets. In general, age and gender recognition from facial images tends to be more reliable than speech alone. Our proposed annotation framework dynamically allocates higher weights to visual agents in such tasks, leading to more accurate and robust annotations. \textcolor{black}{Detailed results for other models and fine-grained metrics are provided in Appendix D.}
\begin{table*}[htbp]
  \centering
  \small         % 或 \footnotesize、\scriptsize
  \caption{Comparison of age and gender estimation metrics.}
  \label{tab:age_gender_metrics}
  \begin{tabular}{lccc}
    \toprule
    \textbf{Method} & \textbf{Age-MAE $\downarrow$} & \textbf{Age-acc(\%) $\uparrow$} & \textbf{Gender-acc(\%) $\uparrow$} \\
    \midrule
    \textbf{Speech Craft~\cite{10.1145/3664647.3681674}}   & 12.54 & 38.02 & 80.69 \\
    \textbf{$\text{M}^3\text{PDB}$(Our)}   & \textbf{6.22}  & \textbf{60.78} & \textbf{91.22} \\
    \bottomrule
  \end{tabular}
  \vspace{-0.3cm}
\end{table*}
\subsection{Candidate Annotation for Unseen Language}
%在通过3.3所提方法获得未见语言的参考数据集后，我们会使用VoxPopuli中的同一语言作为基准测试集。我们会从VoxPopuli选择这个语言的语音，并保持DNSMOS的均值和方差和未见语言的参考数据集一致。
\paragraph{Datasets} 
In this experiment, we treat Italian as an unseen language. Following the method described in Section \ref{sec:3_3}, we construct a candidate set $\text{Prompt}_{can}$ from a different language in our $\text{M}^3\text{PDB}$. Additionally, we extract four language-specific reference sets from the dataset VoxPopuli \cite{wang-etal-2021-voxpopuli}—German $\text{Prompt}_{ge}$, French $\text{Prompt}_{fr}$, Korean $\text{Prompt}_{ko}$, and Italian $\text{Prompt}_{it}$—ensuring that each has the same number of samples and similar audio quality (matched in terms of DNSMOS mean and variance) as $\text{Prompt}_{can}$. These sets are then used as reference audio for Italian speech synthesis. 
\paragraph{Experiment Implementation} 
We utilize XTTS-v2\footnote{https://huggingface.co/coqui/XTTS-v2} as the synthesis model to generate Italian speech, using prompt audio sampled from various languages set. The Italian text are randomly selected from VoxPopuli.

\paragraph{Metrics}
We evaluate the quality of the synthesized Italian speech from the following aspects: (1) the language probability (\textbf{LI}) that the generated audio is classified as Italian; (2) the similarity between the synthesized audio and the reference prompt in terms of speaker (\textbf{SS}), emotion (\textbf{ES}), and speaking rate (\textbf{SRS}); and (3) the content quality of the generated speech, measured by Character Error Rate (\textbf{CER}).
\textcolor{black}{Please refer to the Appendix E for the detailed computation of the evaluation metrics.}
\paragraph{Results}
Table \ref{tab:3} shows the synthesis results of Italian speech using different reference audio sets. First, by comparing the results using reference sets $\text{Prompt}_{can}$ and $\text{Prompt}_{it}$, we observe that using native Italian audio as the reference does not necessarily lead to the highest-quality synthesized speech. In particular, the content quality, as measured by CER, is relatively poor, which may indicate that the Italian reference samples suffer from lower audio quality.
In contrast, when using cross-lingual references $\text{Prompt}_{can}$ selected by our proposed method, the synthesized Italian speech consistently performs better across multiple evaluation metrics. Furthermore, we find that the references from set $\text{Prompt}_{can}$ produce better results than randomly selected audios from other languages. 
Actually, $\text{Prompt}_{can}$ is also French, but it is more suitable for synthesizing Italian speech than for being randomly chosen from French. This highlights the effectiveness of our method.

\begin{table*}[htbp]

  \centering
  \small         % 或 \footnotesize、\scriptsize
  \caption{Comparison of annotation metrics across different languages. All methods except A achieve an error bar within 5\%.}
  \label{tab:language_annotation_metrics}
  \begin{tabular}{lccccc}
    \toprule
    \textbf{Prompt source} & \textbf{LP(\%) $\uparrow$} & \textbf{SS(\%) $\uparrow$} & \textbf{ES(\%) $\uparrow$} & \textbf{CER(\%) $\downarrow$} & \textbf{SRS(\%) $\downarrow$} \\
    \midrule
    \textbf{$\text{Prompt}_{fr}$} & 95.68 & 84.66 & 65.26 & 5.21  & 30.71   \\
    \textbf{$\text{Prompt}_{ge}$} & 96.88 & 83.69 & 58.00 & 5.20  & 28.11   \\
    \textbf{$\text{Prompt}_{ko}$} & 93.32 & 82.97 & 52.94 & 21.38 & 31.74   \\
    \textbf{$\text{Prompt}_{it}$} & 99.81 & 83.69 & 63.40 & 4.32  & 26.30   \\
    \textbf{$\text{Prompt}_{can}$} & 95.66 & 84.70 & 96.40 & 2.78  & 22.70   \\
    \bottomrule
  \end{tabular}
  \label{tab:3}
  \vspace{-0.3cm}
\end{table*}
\subsection{Multi-model Prompt}
\paragraph{Data}
This section uses the same dataset as in Section \ref{sec:dataset1} as test, but additionally includes the text descriptions accompanying to the audio. 
\paragraph{Experiment Implementation}
We compare two prompt selection strategies in the scenario where only visual modality (i.e., facial image) is available as the prompt. 
The first strategy adopts a SOTA face-to-speech model Imaginary Voice \cite{lee2023imaginary}, where the facial image is directly used to synthesize prompt speech.
The second follows our proposed selection method in section \ref{sec:4_1}, which the system selects the most acoustically suitable reference speech from the $\text{M}^3\text{DBP}$ based on the facial similarity. This section evaluates which strategy yields a better speech prompt in terms of the downstream TTS performance. 
\paragraph{Metrics} 
We compare the speaker similarity (SS) between the speech prompts generated by the two selection methods and the original speech. We also evaluate the overall audio quality (UTMOSv2) of the generated speech \cite{baba2024t05}.

\paragraph{Result} As shown in Table \ref{tab:imaginary_vs_proposed}, the audio selected from the dataset $\text{M}^3\text{PDB}$ using our proposed method demonstrates comparable speaker similarity to the audio synthesized by face-to-speech. However, the selected audio from $\text{M}^3\text{PDB}$ clearly exhibits higher speech quality, highlighting the current limitations of the face-to-speech model in generating high-quality speech. Moreover, compared to the synthesized audio from face-to-speech, the speech samples selected from $\text{M}^3\text{PDB}$ show greater variation and cover a wider range of emotional expressions, making them more suitable as reference prompts.
\textcolor{black}{Additional modality results are provided in Appendix F}

\begin{table*}[htbp]
  \centering
  \small
  \caption{Comparison between Imaginary Voice and the proposed method.}
  \label{tab:imaginary_vs_proposed}
  \begin{tabular}{lcc}
    \toprule
    \textbf{Method}       & \textbf{SS(\%) $\uparrow$} & \textbf{UTMOSv2 $\uparrow$} \\
    \midrule
    \textbf{Imaginary Voice~\cite{lee2023imaginary}}        & \textbf{16.18}                          & 2.10            \\
    \textbf{$\text{M}^3\text{PDB}$}          & 15.14                          & \textbf{2.69}           \\
    \bottomrule
  \end{tabular}
  %\vspace{-0.5cm}
\end{table*}

\subsection{Online Selection}

%本部分采用了4.1中通过语音进行注册后，筛选得到的$\text{M}^3\text{PDB}_{small}$数据集作为prompt dataset。需要说明的是，4.1采用何种模态进行注册，并不影响评价4.2中各种方法的指标的优劣关系。使用了4.1的注册音频及同一说话人的其他音频作为Baseline speech generation models 的测试输入，这部分音频额外加入了噪声和混响，从而更接近真实环境。
\paragraph{Data} 
 We randomly selected 20 speakers from the LRS3 \cite{afouras2018lrs3} dataset for evaluation. For each speaker, one audio sample was randomly chosen for registration to construct the reference set $\text{M}^3\text{PDB}$. Additionally, 20 other audio samples were selected per speaker as real-time inputs to evaluate the effectiveness of our online selection strategy. To better simulate real-world conditions, random noise is added to the audio during real-time input. 
\paragraph{Experiment Implementation}
For real-time audio inputs, we compare the following three reference selection strategies: 1) \textbf{Original}: the input audio is used directly as the prompt; 2) \textbf{Random}: a sample is randomly selected from the $\mathrm{M}^3\mathrm{PDB}_{\text{small}}$ dataset to serve as the prompt. 3) \textbf{Proposed}: we adopt our proposed online selection strategy to choose reference audio from set $\mathrm{M}^3\mathrm{PDB}_{\text{small}}$ as the prompt. 
 \textcolor{black}{In this experiment, we constructed four similarity modules in series: the first estimates speech‐rate similarity, the second computes pitch similarity, the third evaluates speaker similarity, and the fourth assesses emotional similarity. The $Top_{i}$ thresholds were set at $Top_1={100\%}$, $Top_2={20\%}$, $Top_3={20\%}$ and $Top_4={20\%}$.}

Furthermore, evaluating online Selection in a streaming setting requires a speech-translation model that supports streaming inference and speaker cloning. Since no existing public model meets these criteria, we build a tailored streaming speech–translation architecture using seamless and cosyvoice2. 
\textcolor{black}{The model’s detail structure are detailed in Appendix G.}
All experiments were conducted on a single NVIDIA A100-PCIE-40GB.

\paragraph{Metrics}
In this section, we evaluate the speaker similarity (SS), emotion similarity (ES), and speaking rate similarity (SRS) between the input and the generated speech in speech-to-speech translation. Additionally, we assess the speech quality of the generated speech using UTMOSv2 and character error rate (CER).
%结果如表1所示，可以看到，所提方法在DNSMOS
\begin{table*}[htbp]
  %\vspace{-0.5cm}
  \centering
  \small
  \caption{Comparison of methods on SS, ES, CER, SRS and DNSMOS metrics.}
  \label{tab:metrics_comparison}
  \begin{tabular}{lccccc}
    \toprule
    \textbf{Method}      & \textbf{SS(\%) $\uparrow$} & \textbf{ES(\%) $\uparrow$} & \textbf{CER(\%) $\downarrow$} & \textbf{SRS(\%) $\downarrow$} & \textbf{UTMOSv2 $\uparrow$} \\
    \midrule
    \textbf{Original input}       & 39.53                & 65.26                & 4.21                    & 30.71                   & 2.49                    \\
    \textbf{Random}               & 3.69                & 18.00                & 1.34                    & 48.11                   & 2.65                    \\
    \textbf{$\text{M}^3\text{PDB}$}                & \textbf{44.88}                & \textbf{72.94}                & \textbf{1.17}                   & \textbf{28.74}                   & \textbf{2.69}     \\
    \bottomrule
  \end{tabular}
  \label{tab:5}
  \vspace{-0.5cm}
\end{table*}

\paragraph{Result} 
As shown in Table \ref{tab:5}, our proposed online prompt selection strategy consistently enables high-quality speech synthesis, even when the input audio is of low quality. Notably, although the selected prompt audio is not from the same speaker as the input, the synthesized speech still achieves high similarity to the target regarding emotion and other speaker-related characteristics. These results demonstrate the effectiveness and robustness of both our database construction and prompt selection approach under challenging real-world conditions.
\textcolor{black}{Please refer to Appendix H for the performance comparison under varying latency conditions.}

\section{Conclusion, Limitations and Future Works}
This work addresses a critical gap in zero-shot speech generation: the lack of robust solutions for real-world prompt variability. We introduce $\text{M}^3\text{PDB}$, a large-scale, multi-modal, multi-label, and multilingual prompt database designed to support prompt selection under realistic and challenging conditions. 
Specifically, the $\text{M}^3\text{PDB}$ dataset includes annotations across 10 labels for both speech and visual modalities, covering 18 languages, with about 15k speakers and a total of 400k hours of data.
Through a novel multi-agent annotation framework and an efficient selection strategy, our $\text{M}^3\text{PDB}$ demonstrates improved performance in settings where prompts are noisy, incomplete, or entirely missing. 
\paragraph{Future Work and Limitations} 1) Although we have designed an efficient prompt selection strategy based on the $\text{M}^3\text{PDB}$ database, building the database itself remains time-consuming. In the future, we plan to explore methods to accelerate data processing and annotation.
2) The current version of the database covers 18 languages. We aim to expand it to include a broader set of languages.
3) So far, our evaluations have focused on two tasks: speech translation and text-to-speech (TTS). In future work, we plan to extend our analysis to other speech-related applications such as interactive speech systems.
\paragraph{Broader Impact} Beyond empirical results, our work aims to shift the speech generation community’s attention toward more practical and deployment-oriented problems, moving beyond the comfort zone of controlled benchmarks. We release our code and dataset to encourage further research in this direction.

\bibliographystyle{IEEEtran} 
\bibliography{ref.bib}

% Generated by IEEEtran.bst, version: 1.14 (2015/08/26)
\begin{thebibliography}{10}
\providecommand{\url}[1]{#1}
\csname url@samestyle\endcsname
\providecommand{\newblock}{\relax}
\providecommand{\bibinfo}[2]{#2}
\providecommand{\BIBentrySTDinterwordspacing}{\spaceskip=0pt\relax}
\providecommand{\BIBentryALTinterwordstretchfactor}{4}
\providecommand{\BIBentryALTinterwordspacing}{\spaceskip=\fontdimen2\font plus
\BIBentryALTinterwordstretchfactor\fontdimen3\font minus
  \fontdimen4\font\relax}
\providecommand{\BIBforeignlanguage}[2]{{%
\expandafter\ifx\csname l@#1\endcsname\relax
\typeout{** WARNING: IEEEtran.bst: No hyphenation pattern has been}%
\typeout{** loaded for the language `#1'. Using the pattern for}%
\typeout{** the default language instead.}%
\else
\language=\csname l@#1\endcsname
\fi
#2}}
\providecommand{\BIBdecl}{\relax}
\BIBdecl

\bibitem{9054535}
E.~Cooper, C.-I. Lai, Y.~Yasuda, F.~Fang, X.~Wang, N.~Chen, and J.~Yamagishi,
  ``{Zero-Shot Multi-Speaker Text-To-Speech with State-Of-The-Art Neural
  Speaker Embeddings},'' in \emph{ICASSP 2020 - 2020 IEEE International
  Conference on Acoustics, Speech and Signal Processing (ICASSP)}, 2020, pp.
  6184--6188.

\bibitem{10.5555/3692070.3692979}
Z.~Ju, Y.~Wang, K.~Shen, X.~Tan, D.~Xin, D.~Yang, Y.~Liu, Y.~Leng, K.~Song,
  S.~Tang, Z.~Wu, T.~Qin, X.-Y. Li, W.~Ye, S.~Zhang, J.~Bian, L.~He, J.~Li, and
  S.~Zhao, ``{NaturalSpeech 3: zero-shot speech synthesis with factorized codec
  and diffusion models},'' in \emph{Proceedings of the 41st International
  Conference on Machine Learning}, ser. ICML'24.\hskip 1em plus 0.5em minus
  0.4em\relax JMLR.org, 2024.

\bibitem{NEURIPS2023_3eaad2a0}
\BIBentryALTinterwordspacing
Y.~A. Li, C.~Han, V.~Raghavan, G.~Mischler, and N.~Mesgarani, ``{StyleTTS 2:
  Towards Human-Level Text-to-Speech through Style Diffusion and Adversarial
  Training with Large Speech Language Models},'' in \emph{Advances in Neural
  Information Processing Systems}, A.~Oh, T.~Naumann, A.~Globerson, K.~Saenko,
  M.~Hardt, and S.~Levine, Eds., vol.~36.\hskip 1em plus 0.5em minus
  0.4em\relax Curran Associates, Inc., 2023, pp. 19\,594--19\,621. [Online].
  Available:
  \url{https://proceedings.neurips.cc/paper_files/paper/2023/file/3eaad2a0b62b5ed7a2e66c2188bb1449-Paper-Conference.pdf}
\BIBentrySTDinterwordspacing

\bibitem{zen19_interspeech}
H.~Zen, V.~Dang, R.~Clark, Y.~Zhang, R.~J. Weiss, Y.~Jia, Z.~Chen, and Y.~Wu,
  ``{LibriTTS: A Corpus Derived from LibriSpeech for Text-to-Speech},'' in
  \emph{Interspeech 2019}, 2019, pp. 1526--1530.

\bibitem{chen21o_interspeech}
G.~Chen, S.~Chai, G.-B. Wang, J.~Du, W.-Q. Zhang, C.~Weng, D.~Su, D.~Povey,
  J.~Trmal, J.~Zhang, M.~Jin, S.~Khudanpur, S.~Watanabe, S.~Zhao, W.~Zou,
  X.~Li, X.~Yao, Y.~Wang, Z.~You, and Z.~Yan, ``{GigaSpeech: An Evolving,
  Multi-Domain ASR Corpus with 10,000 Hours of Transcribed Audio},'' in
  \emph{Interspeech 2021}, 2021, pp. 3670--3674.

\bibitem{pratap20_interspeech}
V.~Pratap, Q.~Xu, A.~Sriram, G.~Synnaeve, and R.~Collobert, ``{MLS: A
  Large-Scale Multilingual Dataset for Speech Research},'' in \emph{Interspeech
  2020}, 2020, pp. 2757--2761.

\bibitem{10.1145/3664647.3681674}
\BIBentryALTinterwordspacing
Z.~Jin, J.~Jia, Q.~Wang, K.~Li, S.~Zhou, S.~Zhou, X.~Qin, and Z.~Wu,
  ``{SpeechCraft: A Fine-Grained Expressive Speech Dataset with Natural
  Language Description},'' in \emph{Proceedings of the 32nd ACM International
  Conference on Multimedia}, ser. MM '24.\hskip 1em plus 0.5em minus
  0.4em\relax New York, NY, USA: Association for Computing Machinery, 2024, p.
  1255–1264. [Online]. Available:
  \url{https://doi.org/10.1145/3664647.3681674}
\BIBentrySTDinterwordspacing

\bibitem{chu2023qwen}
Y.~Chu, J.~Xu, X.~Zhou, Q.~Yang, S.~Zhang, Z.~Yan, C.~Zhou, and J.~Zhou,
  ``{Qwen-audio: Advancing universal audio understanding via unified
  large-scale audio-language models},'' \emph{arXiv preprint arXiv:2311.07919},
  2023.

\bibitem{an2024funaudiollm}
K.~An, Q.~Chen, C.~Deng, Z.~Du, C.~Gao, Z.~Gao, Y.~Gu, T.~He, H.~Hu, K.~Hu
  \emph{et~al.}, ``{Funaudiollm: Voice understanding and generation foundation
  models for natural interaction between humans and llms},'' \emph{arXiv
  preprint arXiv:2407.04051}, 2024.

\bibitem{cheng2024videollama}
Z.~Cheng, S.~Leng, H.~Zhang, Y.~Xin, X.~Li, G.~Chen, Y.~Zhu, W.~Zhang, Z.~Luo,
  D.~Zhao \emph{et~al.}, ``{Videollama 2: Advancing spatial-temporal modeling
  and audio understanding in video-llms},'' \emph{arXiv preprint
  arXiv:2406.07476}, 2024.

\bibitem{ma24d_interspeech}
L.~Ma, D.~Guo, K.~Song, Y.~Jiang, S.~Wang, L.~Xue, W.~Xu, H.~Zhao, B.~Zhang,
  and L.~Xie, ``{WenetSpeech4TTS: A 12,800-hour Mandarin TTS Corpus for Large
  Speech Generation Model Benchmark},'' in \emph{Interspeech 2024}, 2024, pp.
  1840--1844.

\bibitem{he2025emilia}
H.~He, Z.~Shang, C.~Wang, X.~Li, Y.~Gu, H.~Hua, L.~Liu, C.~Yang, J.~Li, P.~Shi
  \emph{et~al.}, ``{Emilia: A Large-Scale, Extensive, Multilingual, and Diverse
  Dataset for Speech Generation},'' \emph{arXiv preprint arXiv:2501.15907},
  2025.

\bibitem{10842513}
S.~Chen, C.~Wang, Y.~Wu, Z.~Zhang, L.~Zhou, S.~Liu, Z.~Chen, Y.~Liu, H.~Wang,
  J.~Li, L.~He, S.~Zhao, and F.~Wei, ``{Neural Codec Language Models are
  Zero-Shot Text to Speech Synthesizers},'' \emph{IEEE Transactions on Audio,
  Speech and Language Processing}, vol.~33, pp. 705--718, 2025.

\bibitem{du2024cosyvoice}
Z.~Du, Y.~Wang, Q.~Chen, X.~Shi, X.~Lv, T.~Zhao, Z.~Gao, Y.~Yang, C.~Gao,
  H.~Wang \emph{et~al.}, ``{Cosyvoice 2: Scalable streaming speech synthesis
  with large language models},'' \emph{arXiv preprint arXiv:2412.10117}, 2024.

\bibitem{anastassiou2024seed}
P.~Anastassiou, J.~Chen, J.~Chen, Y.~Chen, Z.~Chen, Z.~Chen, J.~Cong, L.~Deng,
  C.~Ding, L.~Gao \emph{et~al.}, ``{Seed-tts: A family of high-quality
  versatile speech generation models},'' \emph{arXiv preprint
  arXiv:2406.02430}, 2024.

\bibitem{wang2025spark}
X.~Wang, M.~Jiang, Z.~Ma, Z.~Zhang, S.~Liu, L.~Li, Z.~Liang, Q.~Zheng, R.~Wang,
  X.~Feng \emph{et~al.}, ``{Spark-tts: An efficient llm-based text-to-speech
  model with single-stream decoupled speech tokens},'' \emph{arXiv preprint
  arXiv:2503.01710}, 2025.

\bibitem{chen2024vall}
S.~Chen, S.~Liu, L.~Zhou, Y.~Liu, X.~Tan, J.~Li, S.~Zhao, Y.~Qian, and F.~Wei,
  ``{Vall-e 2: Neural codec language models are human parity zero-shot text to
  speech synthesizers},'' \emph{arXiv preprint arXiv:2406.05370}, 2024.

\bibitem{10094745}
J.~Lee, J.~Son~Chung, and S.-W. Chung, ``{Imaginary Voice: Face-Styled
  Diffusion Model for Text-to-Speech},'' in \emph{ICASSP 2023 - 2023 IEEE
  International Conference on Acoustics, Speech and Signal Processing
  (ICASSP)}, 2023, pp. 1--5.

\bibitem{lee24_interspeech}
M.~Lee, E.~Park, and S.~Hong, ``{FVTTS : Face Based Voice Synthesis for
  Text-to-Speech},'' in \emph{Interspeech 2024}, 2024, pp. 4953--4957.

\bibitem{barrault2023seamless}
L.~Barrault, Y.-A. Chung, M.~C. Meglioli, D.~Dale, N.~Dong, M.~Duppenthaler,
  P.-A. Duquenne, B.~Ellis, H.~Elsahar, J.~Haaheim \emph{et~al.}, ``{Seamless:
  Multilingual Expressive and Streaming Speech Translation},'' \emph{arXiv
  preprint arXiv:2312.05187}, 2023.

\bibitem{NEURIPS2024_a32539cb}
\BIBentryALTinterwordspacing
C.~Le, Y.~Qian, D.~Wang, L.~Zhou, S.~Liu, X.~Wang, M.~Yousefi, Y.~Qian, J.~Li,
  S.~Zhao, and M.~Zeng, ``{TransVIP: Speech to Speech Translation System with
  Voice and Isochrony Preservation},'' in \emph{Advances in Neural Information
  Processing Systems}, A.~Globerson, L.~Mackey, D.~Belgrave, A.~Fan, U.~Paquet,
  J.~Tomczak, and C.~Zhang, Eds., vol.~37.\hskip 1em plus 0.5em minus
  0.4em\relax Curran Associates, Inc., 2024, pp. 89\,682--89\,705. [Online].
  Available:
  \url{https://proceedings.neurips.cc/paper_files/paper/2024/file/a32539cb16274581a17e679f6046f4bf-Paper-Conference.pdf}
\BIBentrySTDinterwordspacing

\bibitem{fu2025vita}
C.~Fu, H.~Lin, X.~Wang, Y.-F. Zhang, Y.~Shen, X.~Liu, H.~Cao, Z.~Long, H.~Gao,
  K.~Li \emph{et~al.}, ``{Vita-1.5: Towards gpt-4o level real-time vision and
  speech interaction},'' \emph{arXiv preprint arXiv:2501.01957}, 2025.

\bibitem{wang2024emopro}
H.~Wang, C.~Qiang, T.~Wang, C.~Gong, Q.~Liu, Y.~Jiang, X.~Wang, C.~Wang, and
  C.~Zhang, ``{EmoPro: A Prompt Selection Strategy for Emotional Expression in
  LM-based Speech Synthesis},'' \emph{arXiv preprint arXiv:2409.18512}, 2024.

\bibitem{9747801}
C.~Gong, L.~Wang, Z.~Ling, J.~Zhang, and J.~Dang, ``{Using Multiple Reference
  Audios and Style Embedding Constraints for Speech Synthesis},'' in
  \emph{ICASSP 2022 - 2022 IEEE International Conference on Acoustics, Speech
  and Signal Processing (ICASSP)}, 2022, pp. 7912--7916.

\bibitem{luo2025autostyle}
D.~Luo, C.~Ma, W.~Li, J.~Wang, W.~Chen, and Z.~Wu, ``{AutoStyle-TTS:
  Retrieval-Augmented Generation based Automatic Style Matching Text-to-Speech
  Synthesis},'' \emph{arXiv preprint arXiv:2504.10309}, 2025.

\bibitem{chung2018voxceleb2}
J.~S. Chung, A.~Nagrani, and A.~Zisserman, ``{Voxceleb2: Deep speaker
  recognition},'' \emph{arXiv preprint arXiv:1806.05622}, 2018.

\bibitem{8373813}
Q.~Cao, L.~Shen, W.~Xie, O.~M. Parkhi, and A.~Zisserman, ``{VGGFace2: A Dataset
  for Recognising Faces across Pose and Age},'' in \emph{2018 13th IEEE
  International Conference on Automatic Face \& Gesture Recognition (FG 2018)},
  2018, pp. 67--74.

\bibitem{ma2023emotion2vec}
Z.~Ma, Z.~Zheng, J.~Ye, J.~Li, Z.~Gao, S.~Zhang, and X.~Chen, ``emotion2vec:
  Self-supervised pre-training for speech emotion representation,'' \emph{arXiv
  preprint arXiv:2312.15185}, 2023.

\bibitem{pmlr-v202-longpre23a}
\BIBentryALTinterwordspacing
S.~Longpre, L.~Hou, T.~Vu, A.~Webson, H.~W. Chung, Y.~Tay, D.~Zhou, Q.~V. Le,
  B.~Zoph, J.~Wei, and A.~Roberts, ``The flan collection: Designing data and
  methods for effective instruction tuning,'' in \emph{Proceedings of the 40th
  International Conference on Machine Learning}, ser. Proceedings of Machine
  Learning Research, A.~Krause, E.~Brunskill, K.~Cho, B.~Engelhardt, S.~Sabato,
  and J.~Scarlett, Eds., vol. 202.\hskip 1em plus 0.5em minus 0.4em\relax PMLR,
  23--29 Jul 2023, pp. 22\,631--22\,648. [Online]. Available:
  \url{https://proceedings.mlr.press/v202/longpre23a.html}
\BIBentrySTDinterwordspacing

\bibitem{wang-etal-2021-voxpopuli}
\BIBentryALTinterwordspacing
C.~Wang, M.~Riviere, A.~Lee, A.~Wu, C.~Talnikar, D.~Haziza, M.~Williamson,
  J.~Pino, and E.~Dupoux, ``{V}ox{P}opuli: A large-scale multilingual speech
  corpus for representation learning, semi-supervised learning and
  interpretation,'' in \emph{Proceedings of the 59th Annual Meeting of the
  Association for Computational Linguistics and the 11th International Joint
  Conference on Natural Language Processing (Volume 1: Long Papers)}, C.~Zong,
  F.~Xia, W.~Li, and R.~Navigli, Eds.\hskip 1em plus 0.5em minus 0.4em\relax
  Online: Association for Computational Linguistics, Aug. 2021, pp. 993--1003.
  [Online]. Available: \url{https://aclanthology.org/2021.acl-long.80/}
\BIBentrySTDinterwordspacing

\bibitem{lee2023imaginary}
J.~Lee, J.~S. Chung, and S.-W. Chung, ``{Imaginary voice: Face-styled diffusion
  model for text-to-speech},'' in \emph{ICASSP 2023-2023 IEEE International
  Conference on Acoustics, Speech and Signal Processing (ICASSP)}.\hskip 1em
  plus 0.5em minus 0.4em\relax IEEE, 2023, pp. 1--5.

\bibitem{baba2024t05}
K.~Baba, W.~Nakata, Y.~Saito, and H.~Saruwatari, ``{The T05 system for the
  VoiceMOS challenge 2024: Transfer learning from deep image classifier to
  naturalness MOS prediction of high-quality synthetic speech},'' in \emph{2024
  IEEE Spoken Language Technology Workshop (SLT)}.\hskip 1em plus 0.5em minus
  0.4em\relax IEEE, 2024, pp. 818--824.

\bibitem{afouras2018lrs3}
T.~Afouras, J.~S. Chung, and A.~Zisserman, ``{LRS3-TED: a large-scale dataset
  for visual speech recognition},'' \emph{arXiv preprint arXiv:1809.00496},
  2018.

\bibitem{radford2023robust}
A.~Radford, J.~W. Kim, T.~Xu, G.~Brockman, C.~McLeavey, and I.~Sutskever,
  ``Robust speech recognition via large-scale weak supervision,'' in
  \emph{International conference on machine learning}.\hskip 1em plus 0.5em
  minus 0.4em\relax PMLR, 2023, pp. 28\,492--28\,518.

\bibitem{gong24c_interspeech}
C.~Gong, E.~Cooper, X.~Wang, C.~Qiang, M.~Geng, D.~Wells, L.~Wang, J.~Dang,
  M.~Tessier, A.~Pine, K.~Richmond, and J.~Yamagishi, ``An initial
  investigation of language adaptation for tts systems under low-resource
  scenarios,'' in \emph{Interspeech 2024}, 2024, pp. 4963--4967.

\bibitem{do23_ssw}
P.~Do, M.~Coler, J.~Dijkstra, and E.~Klabbers, ``Strategies in transfer
  learning for low-resource speech synthesis: Phone mapping, features input,
  and source language selection,'' in \emph{12th ISCA Speech Synthesis Workshop
  (SSW2023)}, 2023, pp. 21--26.

\bibitem{guan2024mm}
W.~Guan, Y.~Li, T.~Li, H.~Huang, F.~Wang, J.~Lin, L.~Huang, L.~Li, and Q.~Hong,
  ``Mm-tts: Multi-modal prompt based style transfer for expressive
  text-to-speech synthesis,'' in \emph{Proceedings of the AAAI Conference on
  Artificial Intelligence}, vol.~38, no.~16, 2024, pp. 18\,117--18\,125.

\end{thebibliography}

%%%%%%%%%%%%%%%%%%%%%%%%%%%%%%%%%%%%%%%%%%%%%%%%%%%%%%%%%%%%

\appendix

\section{Technical Appendices and Supplementary Material}

\subsection{Data Distribution and Preprocess}
\begin{figure}[h]
  \centering
  %\vspace{-0.65cm}
  \includegraphics[width=0.9\textwidth]{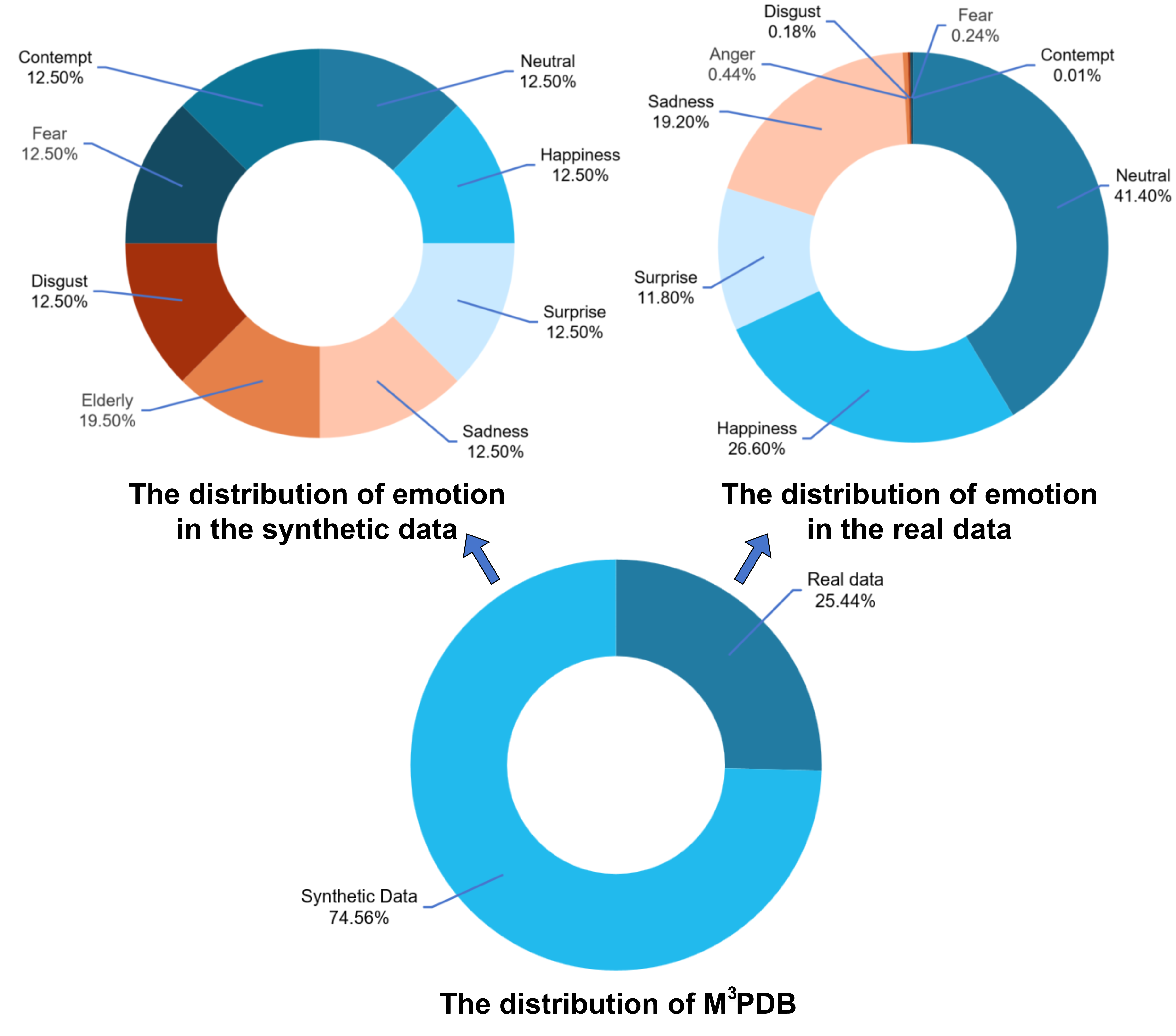}
  %\vspace{-0.9cm}
  \caption{The distribution of $\text{M}^3\text{PDB}$ dataset.}
  \label{figA.1}
  %\vspace{-0.5cm}
  
\end{figure}

\begin{figure}[h]
  \centering
  %\vspace{-0.65cm}
  \includegraphics[width=0.5\textwidth]{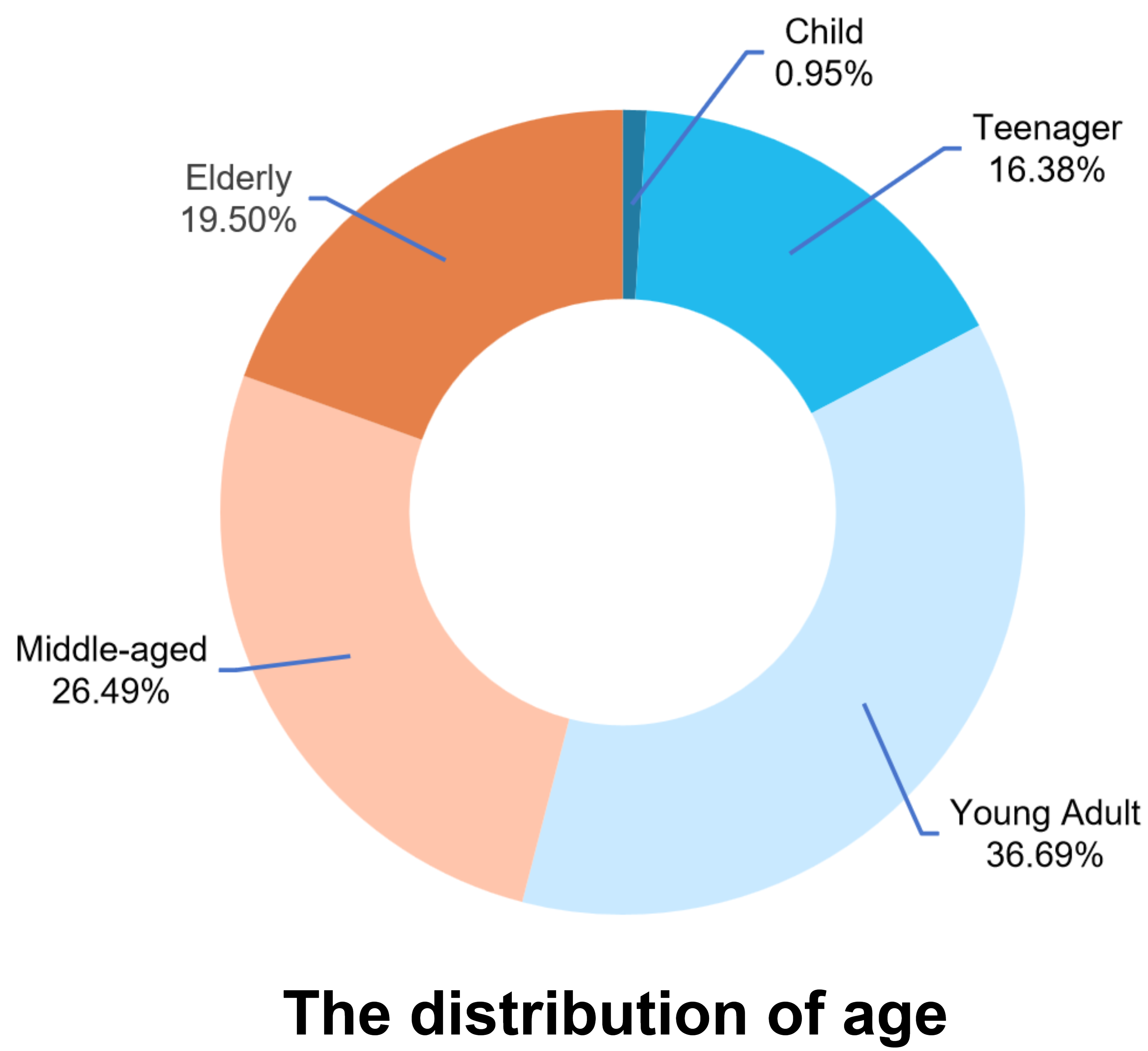}
  %\vspace{-0.9cm}
  \caption{The distribution of age.}
  \label{figA.2}
  %\vspace{-0.5cm}
  
\end{figure}

As shown in Figure \ref{figA.1}, the $\text{M}^3\text{PDB}$ dataset includes both real and synthetic data, with approximately one-fourth of the data being real. This real data is sourced from Emilia, VGG-Sound, VoxCeleb, and VoxPopuli. 
However, the real data show limited variability; approximately half of them express a neutral emotion. In contrast, the synthetic data provides a broader range of emotional categories.
The language distribution is presented in Table \ref{tab:language_hours}. $\text{M}^3\text{PDB}$ covers more than 18 languages, enabling its use in many countries worldwide.
In comparison to other datasets, $\text{M}^3\text{PDB}$ offers a broader range of modalities; about 25\% of the data comprises at least three modalities: video, text, and speech.
The gender distribution is nearly balanced, with slight variations among different age groups, as shown in Figure \ref{figA.2}. The main age categories include Elderly, Teenager, Young Adult, and Middle-aged, with a smaller portion of children.

\begin{table*}[htbp]
  \centering
  \small
  \caption{Language distribution}
  \label{tab:language_hours}
  \begin{tabular}{lcr}
    \toprule
    \textbf{Language}  & \textbf{Proportion (\%)} & \textbf{Hours} \\
    \midrule
    English            & 27.27                & 121,623        \\
    Chinese            & 24.50               & 109,357        \\
    Spanish            &  2.94                &  13,135        \\
    French             &  2.70                &  12,037        \\
    German             &  3.02                &  13,471        \\
    Italian            &  2.89               &  12,913        \\
    Portuguese         &  3.14                &  14,014        \\
    Polish             &  2.81                &  12,515        \\
    Turkish            &  2.92               &  13,015        \\
    Russian            &  2.69                &  12,015        \\
    Dutch              &  3.14                &  14,015        \\
    Czech              &  3.11               &  13,862        \\
    Arabic             &  2.86               &  12,759        \\
    Japanese           &  3.79               &  16,905        \\
    Hungarian          &  2.99               &  13,349        \\
    Korean             &  3.21               &  14,335        \\
    Hindi              &  2.91               &  13,003        \\
    Cantonese          &  3.12               &  13,916        \\
    \bottomrule
  \end{tabular}
\end{table*}

\begin{comment}

%数据分布如图1所示。该数据集与其他数据集的分布最大的区别在于语种和情感的分布。现有的大规模数据集里的语种和情感的分布都极度不均，这限制了其在语音合成中的应用。为了解决该问题，我们采用了多个语音合成模型来合成较为稀少的语种和情感的数据，使得稀有语种和情感也能得到尽可能好的复刻。du2024cosyvoice
As shown in Figure \ref{figA.1}, our dataset differs most markedly from others in its language and emotion distributions. Existing large-scale corpora exhibit extreme skew across both axes, limiting their utility in speech synthesis. To remedy this, we employ multiple TTS models to generate data for under-represented languages and emotions, ensuring even rare cases are faithfully reproduced \cite{du2024cosyvoice,casanova2024xtts}.
%数据预处理的流程如图2所示。预处理的数据包括了音频和音视频。如果是音频，则使用Emilia中的处理办法来进行数据预处理。如果是音视频，则先将其分离为音频和视频，并分别进行格式标准化和增强。然后采用多模态的说话人分离方法，并根据音频的质量指标进行筛选，得到多个音频-视频对。对于数据对中的语音，会采用ASR model来得到转录文本。对于数据对中的视频，则会依次进行脸部分割和正脸检测，得到最终的视频和图片数据。预处理后的数据包括大量的数据对。每个数据对中包括了说话人的语音，转录文本，视频和正脸图片。
\end{comment}

\begin{figure}[h]
  \centering
  %\vspace{-0.65cm}
  \includegraphics[width=1\textwidth]{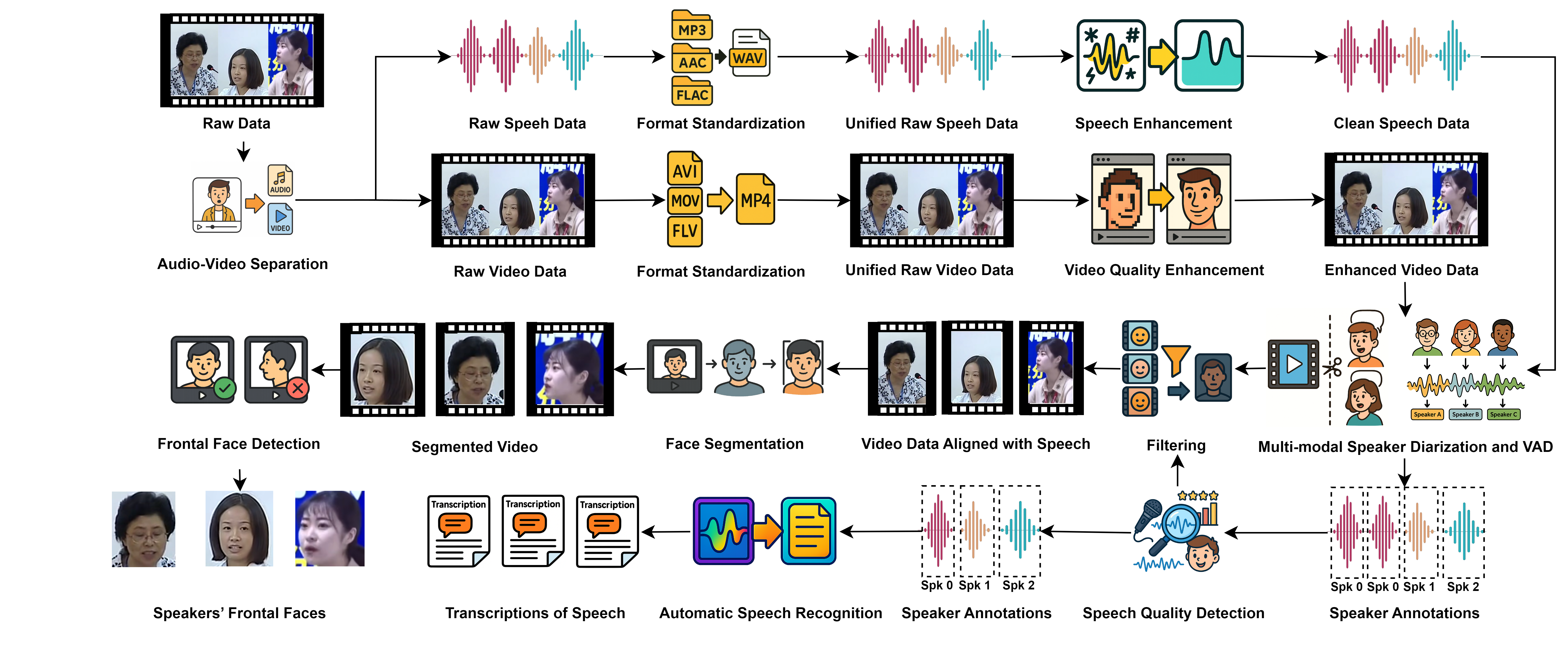}
  %\vspace{-0.9cm}
  \caption{The overall framework for data collection and annotation.}
  \label{figA.3}
  %\vspace{-0.5cm}
\end{figure}

As shown in Figure \ref{figA.3}, our preprocessing pipeline is designed to accommodate both audio-only and audio-video inputs.
For audio and video pairs, we first separate them into individual streams, then standardize formats and apply data augmentation.
Audio files are standardized using the Emilia toolkit\footnote{https://github.com/open-mmlab/Amphion/tree/main/preprocessors/Emilia}.
To improve the accuracy of speaker separation, a multimodal speaker separation model\footnote{https://github.com/modelscope/3D-Speaker} is adopted. 
Audio and video samples of poor quality are filtered out. Each audio segment is then transcribed using an ASR \cite{radford2023robust}, while each video segment undergoes face-mask segmentation followed by frontal face detection\footnote{https://github.com/nawafalageel/Side-Profile-Detection} to create refined video clips and still images.
Finally, the preprocessed data consists of high-quality audio-visual samples, all of which include text transcriptions.
%Next, a multimodal speaker separation model isolates individual speakers, and we filter out segments based on audio-quality metrics to yield clean audio–video pairs. 
%The outcome is a large set of data pairs—each consisting of speaker audio, its transcript, the corresponding video clip, and a canonical face image—for downstream tasks.

\subsection{Configurations of Multi-modal Agent}
%该部分所采用的模型按模态进行了分类，如表1所示。所有使用的模型皆为开源模型，包括了一些学术论文和开源项目中设计的模型。
During the annotation process, we utilized a diverse set of models, as summarized in Table \label{tab:agents}. One of these is a large language model (LLM), which acts as a control unit, providing weights for various agent models. 
Some models are agents that operate across various modalities and annotate multiple attributes of audio-visual data, including age, emotion, gender, language, and more.

%The models used in this section are organized by modality, as shown in Table \ref{tab:1}. All are open-source implementations drawn from both academic publications \cite{ma2024emotion2vec,pmlr-v202-longpre23a,radford2023robust,} and community-driven projects \footnote{https://openai.com/index/gpt-4o-system-card/}\footnote{https://github.com/serengil/deepface}\footnote{https://github.com/librosa/librosa}\footnote{https://huggingface.co/alefiury/wav2vec2-large-xlsr-53-gender-recognition-librispeech}\footnote{https://huggingface.co/audeering/wav2vec2-large-robust-24-ft-age-gender}.

\begin{table*}[htbp]
  \centering
  \small
  \caption{Overview of all models used in multimodal annotation.}
  \label{tab:agents}
  \begin{tabular}{lcc}
    \toprule
    \textbf{Model / Method}          & \textbf{Input Modality}  & \textbf{Output Tag Types}                                     \\
    \midrule
    GPT-4o (RAG)                     & Text                     & Model selection \& weighting                                  \\
    DeepFace                         & Image                    & Gender, Age, Emotion                                          \\
    emotion2vec                      & Audio                    & Emotion                                                       \\
    FLAN-T5                          & Text                     & Emotion                                                       \\
    Whisper                          & Audio                    & Language                                                      \\
    Librosa                        & Audio                    & Speech rate                                                   \\
    MFGCN                            & Audio, Text              & Emotion                                                       \\
    Wav2Vec2-finetuned-age           & Audio                    & Age                                                           \\
    Wav2Vec2-finetuned-gender        & Audio                    & Gender                                                        \\
    OSUM                             & Audio                    & Gender, Age, Emotion, Speech rate                             \\
    SenseVoice                       & Audio                    & Gender, Age, Emotion, Speech rate                             \\
    Qwen2-Audio                      & Audio, Text              & Gender, Age, Emotion, Speech rate, Language, Accent           \\
    \bottomrule
  \end{tabular}
  \vspace{-0.3cm}
\end{table*}
\subsection{Candidate Annotation for Unseen Language}
Figure \ref{} illustrates the detailed annotation process for unseen languages. 
We first compute the language family tree distance between the target language and existing ones in $\text{M}^3\text{PDB}$ based on the linguistic similarity, as described in \cite{gong24c_interspeech,do23_ssw}. 
Next, audio samples from the closest language are further filtered using multiple criteria, the details of which are provided in Section \ref{EM}.

%整个过程的具体过程如图3所示。包括了通过language family tree寻找与未见语种最相近语种，在最相近语种中初步寻找适合该语种的参考语音，以及通过一系列指标来构建最终的未见语种参考数据库。
%Figure 3 presents the complete workflow, leveraging the language family tree to pinpoint the closest known language to an unseen target, harvesting suitable reference utterances from that language, and applying a suite of objective metrics to construct the final reference database for the unseen language.
%\subsection{Closest-Language query method via language family tree}
%\subsection{Closest-Language query method via language family tree}
\subsection{Emotion Annotation Accuracy}
%其他标签的标注准确率的对比结果如表1所示，测试的数据集为IEMOCAP.对于语速计算和语种识别而言，sota的单模态模型已经达到了非常高的acc，因此在这里不额外列出。可以看到，本文所提出方法在情感识别准确率上要显著高于已有的sota模型。
%Table \ref{tab:2} reports comparative annotation accuracies for the remaining labels on the IEMOCAP dataset; since SOTA unimodal models already achieve very high accuracy in speech‐rate estimation and language identification, their results are omitted here, and one can clearly see that our proposed method significantly outperforms existing SOTA models in emotion‐recognition accuracy.
% As shown in Table \ref{tab:2}, annotation accuracies for the other labels are compared; since SOTA unimodal models already achieve very high accuracy in speech‐rate estimation and language identification, those results are omitted here, and our method demonstrates a significant improvement in emotion‐recognition accuracy over existing SOTA models.
In addition to evaluating annotation accuracy on attributes such as age and gender, we also analyze the accuracy of emotion annotations. The evaluation is conducted on the IEMOCAP dataset, and our results are compared against several state-of-the-art (SOTA) emotion prediction models. Experimental results show that our annotation method outperforms both domain-specific emotion models, such as EmoBox, and general large speech models SenseVoice\footnote{}.

\begin{table*}[htbp]
  \centering
  \small
  \caption{Results for emotion annotation accuracy}
  \label{tab:emo}
  \begin{tabular}{lccc}
    \toprule
    \textbf{Method} & \textbf{WA (\%) $\uparrow$} & \textbf{UA (\%) $\uparrow$} & \textbf{F1 (\%) $\uparrow$} \\
    \midrule
    \textbf{SpeechCraft}       & 69.6 & 71.9 & 70.5 \\
    \textbf{EmoBox}            & 72.9 & 73.5 & 73.1 \\
    \textbf{SenseVoice-L}      & 75.3 & 73.9 & 73.2 \\
    \textbf{SenseVoice-S}      & 65.7 & 70.5 & 67.9 \\
    \textbf{$\text{M}^3\text{PDB}$ (Ours)} & \textbf{78.2} & \textbf{77.3} & \textbf{78.3} \\
    \bottomrule
  \end{tabular}
  %\vspace{-0.5cm}
\end{table*}
\subsection{Evaluation Metrics}
\label{EM}
%该部分的指标计算方法与appendix C的指标计算方法基本一致，但在采用的模型和计算细节上有一定区别。区别在于SRS的语速计算是基于whisper识别后的文本除以时长后进行计算的
%LI的计算采用了whipser的语种识别功能，输出了前一层的语种概率，作为language probability。
%SS的计算采用了开源声纹算法储存库里的声纹识别方法，输出了前一层的说话人向量余弦相似度，作为SS。
%ES的计算采用了emotion2vec提取得到的情绪向量，对两个向量计算余弦相似度，作为ES。
%SRS的计算基于以下的计算方法：
%SRS = (speed_test-speed_ref)/speed_ref
%其中speed_test和speed_ref分别为通过librosa测量节拍得到的测试语音和参考语音的语速。
% Language Identification (LI)  
% LI is computed by extracting the language‐probability vector from the penultimate layer of Whisper.

% % Speaker Similarity (SS)  
% SS is defined as the cosine similarity between speaker embeddings obtained from an open‐source speaker‐recognition library.

% % Emotion Similarity (ES)  
% ES is calculated as the cosine similarity of emotion vectors extracted by Emotion2Vec.

% % Speech‐Rate Shift (SRS)  
% \[
% \mathrm{SRS} = \frac{\mathrm{speed}_{\mathrm{test}} - \mathrm{speed}_{\mathrm{ref}}}{\mathrm{speed}_{\mathrm{ref}}},
% \]
% where \(\mathrm{speed}_{\mathrm{test}}\) and \(\mathrm{speed}_{\mathrm{ref}}\) are the speech rates of the test and reference utterances, respectively, as measured by Librosa.
The method for calculating the evaluation metrics is as follows:
\begin{itemize}[leftmargin=*]
    \item \textbf{LI:} LI is computed by extracting the language‐probability vector from the penultimate layer of Whisper.
    \item \textbf{SS:} SS is defined as the cosine similarity between speaker embeddings obtained from an open‐source speaker‐recognition library\footnote{}. 
    \item \textbf{ES:} ES is calculated as the cosine similarity of emotion vectors extracted by Emotion2Vec\footnote{}.
    \item \textbf{SRS:} SRS is calculated as follows:
    \begin{equation}
    \mathrm{SRS} \;=\; \left|\frac{\mathrm{speed}_{\mathrm{test}} - \mathrm{speed}_{\mathrm{ref}}}{\mathrm{speed}_{\mathrm{ref}}}\right|
    \end{equation}
    where \(\mathrm{speed}_{\mathrm{test}}\) and \(\mathrm{speed}_{\mathrm{ref}}\) are the speech rates of the test and reference speech, respectively, as measured by Librosa.
    \item \textbf{CER: }CER is computed by aligning the Whisper‐generated transcript against the ground truth. It is then calculated as follows:
%根据whisper识别得到的文本和groundtruth的文本来计算CER.% Character Error Rate (CER) formula的计算如下所示：
    \begin{equation}
    \mathrm{CER} \;=\; \frac{S + D + I}{N}
    \end{equation}
    
    where \(S\), \(D\), and \(I\) denote the numbers of substitutions, deletions, and insertions respectively, and \(N\) is the total number of characters in the reference transcript.

\end{itemize}
\subsection{Experimental Results of Other Multi-model Prompt Selection}
%其他模态的跨模态选取的对比实验结果如表2所示。可以看到，基于文本或者语音的选取结果同样可以获得较好的语音合成质量。基于语音的匹配结果采用了已有开源模型，与一般声纹识别并无区别，所以在这里不做额外对比实验。
%Table \ref{tab:3} presents comparative results for cross-modal selection in other modalities: both text-driven and audio-driven reference choices deliver similarly high synthesis quality, and because our audio-based matching relies on off-the-shelf, open-source speaker-recognition models identical to the standard approach, no further comparisons are included. 
While the main text has already demonstrated the effectiveness of visual prompts, this section focuses on validating the effectiveness of text-based prompting. 
We use models MM-StyleSpeech and MM-TTS as baselines and evaluate them on the same test set. 
Given text prompts, we compare the speech generated by MM-StyleSpeech and MM-TTS with speech samples selected from $\text{M}^3\text{PDB}$ using the method described in Section 4.1. 
We evaluate the accuracy of the age and gender attributes in the generated and selected prompt speech and the overall audio quality.
As shown in Table \ref{textp}, the audio selected using the method described in Section 4.1 achieves higher accuracy in both gender and age attributes compared to the synthesized audio. Moreover, the audio selected from $\text{M}^3\text{PDB}$ exhibits better audio quality, making it more suitable as a prompt for generation.

\begin{table*}[htbp]
  \centering
  \small         % 或 \footnotesize、\scriptsize
  \caption{Results of text prompt.}
  \label{textp}
  \begin{tabular}{lccc}
    \toprule
    \textbf{Method} & \textbf{Age-acc(\%) $\uparrow$} & \textbf{Gender-acc(\%) $\uparrow$} & \textbf{UTMOSv2 $\uparrow$}\\
    \midrule
    \textbf{MM-StyleSpeech~\cite{guan2024mm}}   & N/A & 98.9 & 1.86 \\
    \textbf{MM-TTS~\cite{guan2024mm}}   & N/A & \textbf{99.6} & 1.83 \\
    \textbf{$\text{M}^3\text{PDB}$(Our)}   & \textbf{60.78} & 91.22 & \textbf{2.69} \\
    \bottomrule
  \end{tabular}
  \vspace{-0.3cm}
\end{table*}

\subsection{Structure of Streaming Translation Model}
\begin{figure}[]
  \centering
  %\vspace{-0.65cm}
  \includegraphics[width=0.9\textwidth]{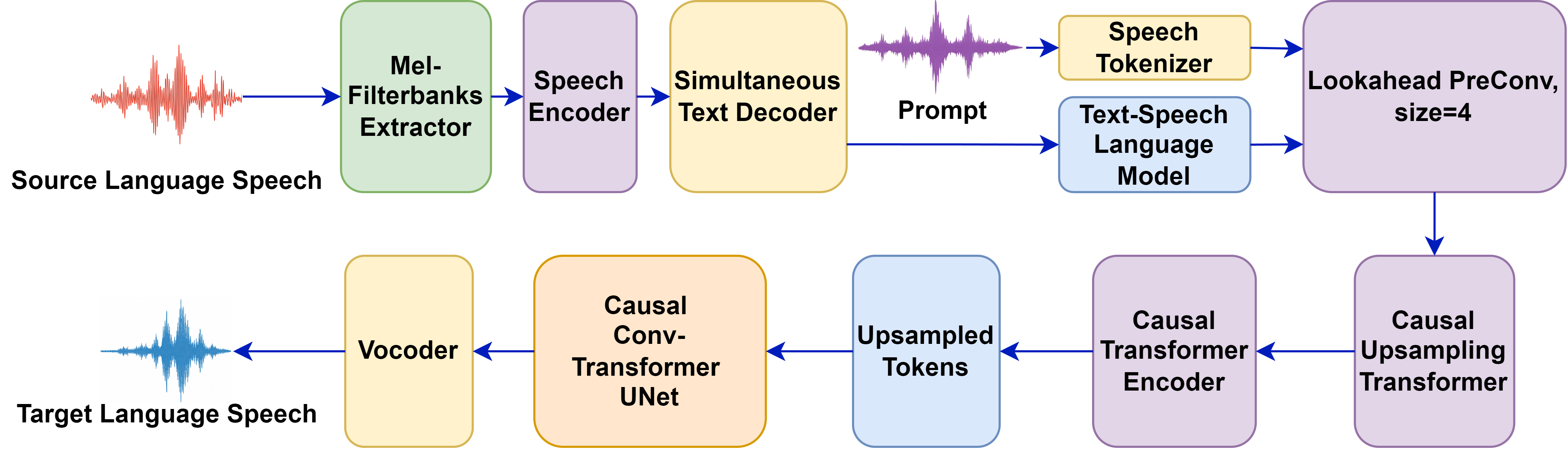}
  %\vspace{-0.9cm}
  \caption{Streaming speech-to-speech translation model.}
  \label{figG}
  %\vspace{-0.5cm}
  
\end{figure}

%本文所提出的流式翻译模型的结构图如图1所示。该模型采用了seamless streaming和cosyvocie2的部分结构，实现了全流程的流式推理，相对于已有的开源模型而言，可以获得更短的推理时间延迟。
Figure \ref{figG} illustrates the architecture of streaming translation model we built, which integrates Seamless-streaming \footnote{} and CosyVoice 2\footnote{}.
Seamless acts as the translation component, enabling streaming speech input and generating streaming text output. Cosyvoice2 serves as the synthesis component, transforming input text into speech with streaming output. During synthesis, the reference audio is no longer restricted to the input audio stream; instead, it can be replaced with high-quality audio from $\text{M}^3\text{PDB}$.

%components to enable end-to-end streaming inference and achieves lower latency than existing open-source solutions.

\subsection{Performance under Varying Latency Conditions}
Our proposed selection strategy supports interruption at any time, allowing the system to return the currently optimal audio selected from $\text{M}^3\text{PDB}$. As described in the main text Section , the selection process considers multiple modules with different $Top_k$.

\begin{figure}[]
  \centering
  %\vspace{-0.65cm}
  \includegraphics[width=0.6\textwidth]{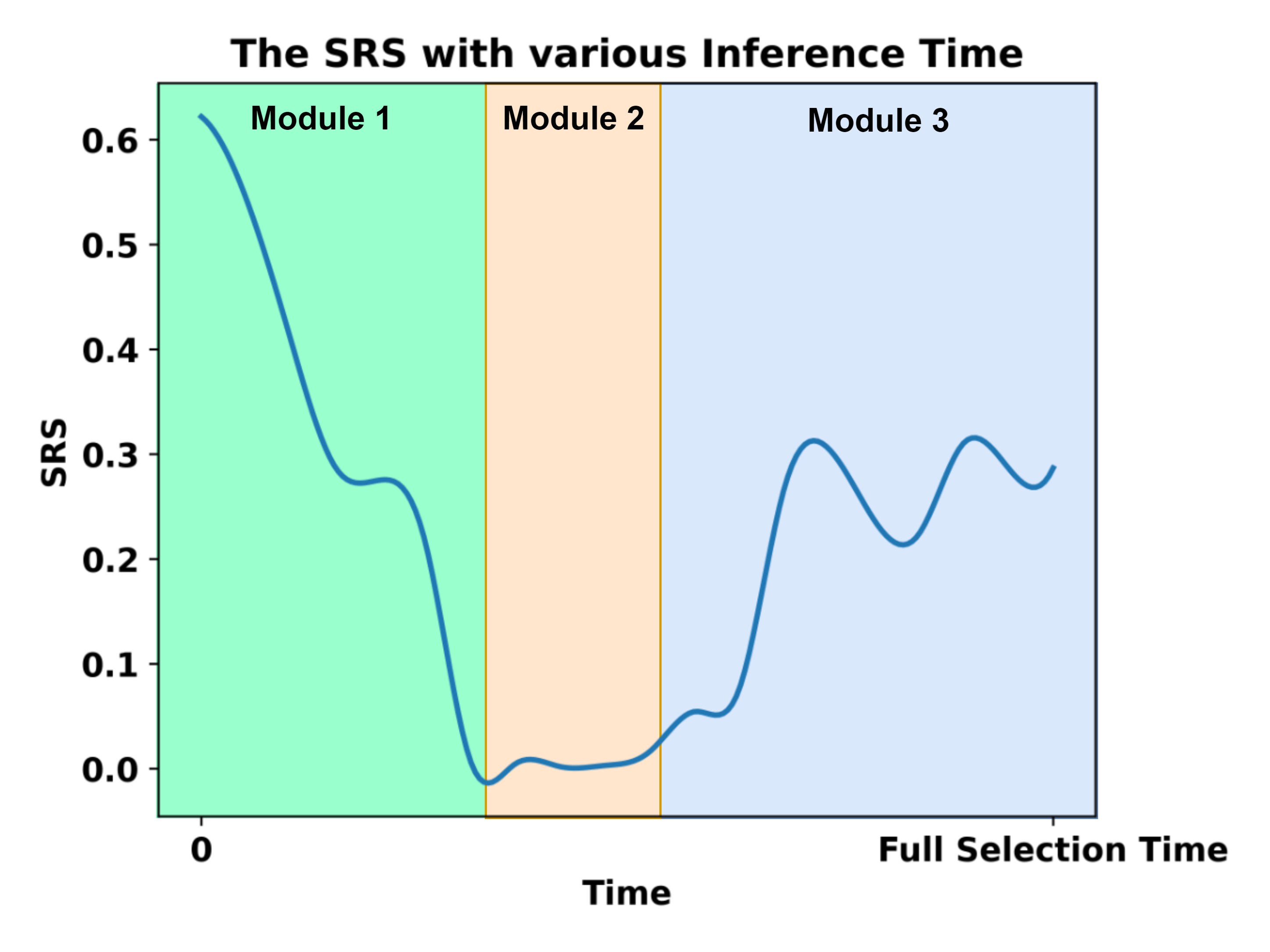}
  %\vspace{-0.9cm}
  \caption{Streaming speech-to-speech translation model.}
  \label{figH1}
  %\vspace{-0.5cm}
  
\end{figure}

\begin{figure}[]
  \centering
  %\vspace{-0.65cm}
  \includegraphics[width=0.6\textwidth]{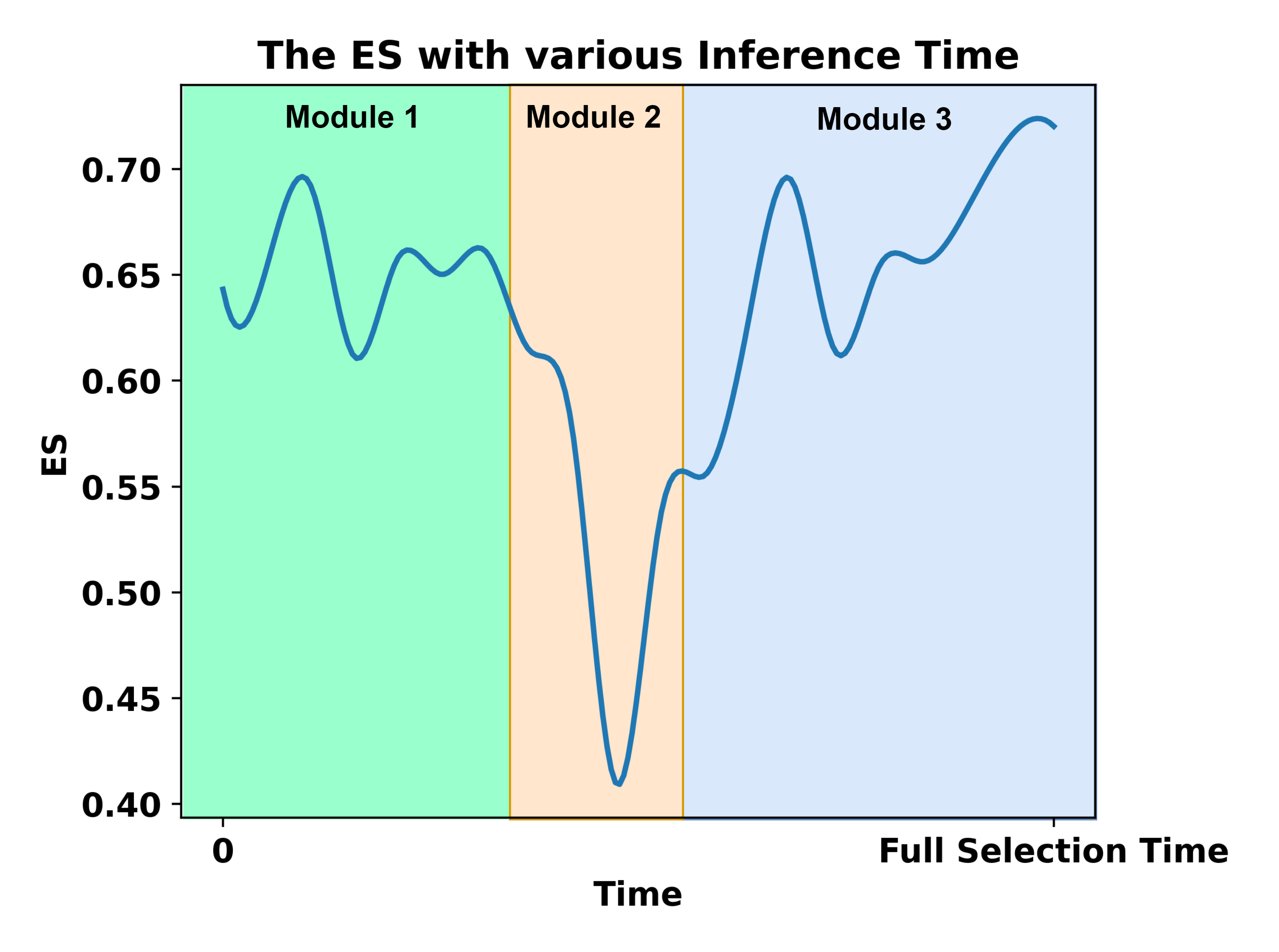}
  %\vspace{-0.9cm}
  \caption{Streaming speech-to-speech translation model.}
  \label{figH2}
  %\vspace{-0.5cm}
  
\end{figure}

\begin{figure}[]
  \centering
  %\vspace{-0.65cm}
  \includegraphics[width=0.6\textwidth]{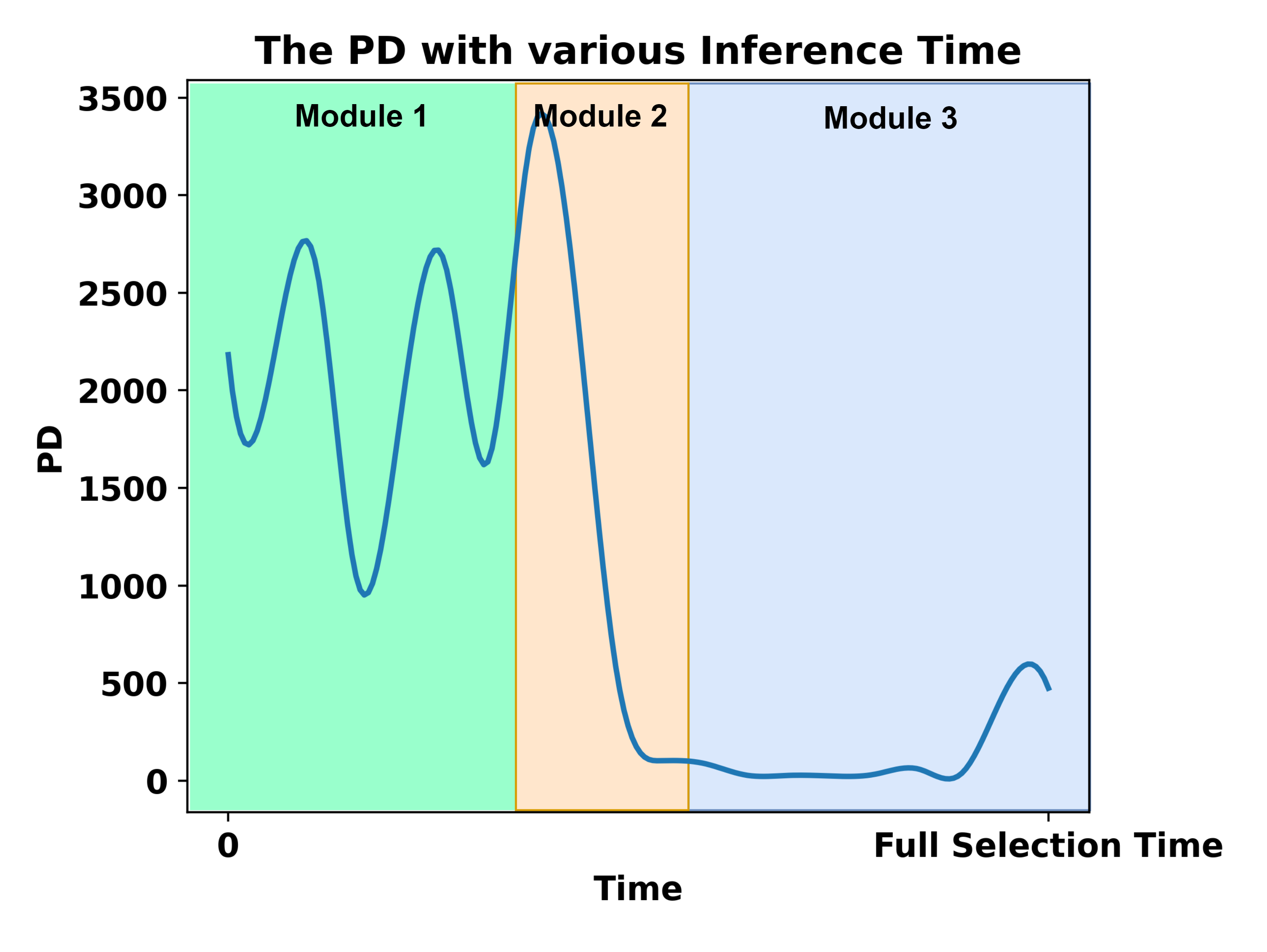}
  %\vspace{-0.9cm}
  \caption{Streaming speech-to-speech translation model.}
  \label{figH3}
  %\vspace{-0.5cm}
  
\end{figure}

As shown in Figures \ref{figH1}, \ref{figH2}, and \ref{figH3}, the quality of the selected audio varies depending on the interruption time. For example, if speaking rate is prioritized, early interruption tends to yield audio with higher similarity. In contrast, if emotion alignment is the focus, later interruption produces more emotionally consistent results. This behavior is closely related to the execution order of different processing modules. In practical usage, users can flexibly adjust the interruption point and module order to balance latency and the desired audio characteristics.

%%%%%%%%%%%%%%%%%%%%%%%%%%%%%%%%%%%%%%%%%%%%%%%%%%%%%%%%%%%%

\newpage
\section*{NeurIPS Paper Checklist}

%%% BEGIN INSTRUCTIONS %%%
The checklist is designed to encourage best practices for responsible machine learning research, addressing issues of reproducibility, transparency, research ethics, and societal impact. Do not remove the checklist: {\bf The papers not including the checklist will be desk rejected.} The checklist should follow the references and follow the (optional) supplemental material.  The checklist does NOT count towards the page
limit. 

Please read the checklist guidelines carefully for information on how to answer these questions. For each question in the checklist:
\begin{itemize}
    \item You should answer \answerYes{}, \answerNo{}, or \answerNA{}.
    \item \answerNA{} means either that the question is Not Applicable for that particular paper or the relevant information is Not Available.
    \item Please provide a short (1–2 sentence) justification right after your answer (even for NA). 
   % \item {\bf The papers not including the checklist will be desk rejected.}
\end{itemize}

{\bf The checklist answers are an integral part of your paper submission.} They are visible to the reviewers, area chairs, senior area chairs, and ethics reviewers. You will be asked to also include it (after eventual revisions) with the final version of your paper, and its final version will be published with the paper.

The reviewers of your paper will be asked to use the checklist as one of the factors in their evaluation. While "\answerYes{}" is generally preferable to "\answerNo{}", it is perfectly acceptable to answer "\answerNo{}" provided a proper justification is given (e.g., "error bars are not reported because it would be too computationally expensive" or "we were unable to find the license for the dataset we used"). In general, answering "\answerNo{}" or "\answerNA{}" is not grounds for rejection. While the questions are phrased in a binary way, we acknowledge that the true answer is often more nuanced, so please just use your best judgment and write a justification to elaborate. All supporting evidence can appear either in the main paper or the supplemental material, provided in appendix. If you answer \answerYes{} to a question, in the justification please point to the section(s) where related material for the question can be found.

IMPORTANT, please:
\begin{itemize}
    \item {\bf Delete this instruction block, but keep the section heading ``NeurIPS Paper Checklist"},
    \item  {\bf Keep the checklist subsection headings, questions/answers and guidelines below.}
    \item {\bf Do not modify the questions and only use the provided macros for your answers}.
\end{itemize}

%%% END INSTRUCTIONS %%%

\begin{enumerate}

\item {\bf Claims}
    \item[] Question: Do the main claims made in the abstract and introduction accurately reflect the paper's contributions and scope?
    \item[] Answer: \answerYes{} % Replace by \answerYes{}, \answerNo{}, or \answerNA{}.
    \item[] Justification: We clearly state our scope and contribution in the abstract and the introduction.

\item {\bf Limitations}
    \item[] Question: Does the paper discuss the limitations of the work performed by the authors?
    \item[] Answer: \answerYes{} % Replace by \answerYes{}, \answerNo{}, or \answerNA{}.
    \item[] Justification: We discussed the limitation of the work in the conclusion section.

\item {\bf Theory assumptions and proofs}
    \item[] Question: For each theoretical result, does the paper provide the full set of assumptions and a complete (and correct) proof?
    \item[] Answer: \answerNA{} % Replace by \answerYes{}, \answerNo{}, or \answerNA{}.
    \item[] Justification: This paper does not include sophisticated experiment result.

    \item {\bf Experimental result reproducibility}
    \item[] Question: Does the paper fully disclose all the information needed to reproduce the main experimental results of the paper to the extent that it affects the main claims and/or conclusions of the paper (regardless of whether the code and data are provided or not)?
    \item[] Answer: \answerYes{} % Replace by \answerYes{}, \answerNo{}, or \answerNA{}.
    \item[] Justification: We believe this paper has contained enough information to reproduce the main result.We provide details about model structure, data processing, training, inference and evaluation.

\item {\bf Open access to data and code}
    \item[] Question: Does the paper provide open access to the data and code, with sufficient instructions to faithfully reproduce the main experimental results, as described in supplemental material?
    \item[] Answer: \answerYes{} % Replace by \answerYes{}, \answerNo{}, or \answerNA{}.
    \item[] Justification: We include the code and the data preparation script in the supplementary material. And we also include open source github url.

\item {\bf Experimental setting/details}
    \item[] Question: Does the paper specify all the training and test details (e.g., data splits, hyperparameters, how they were chosen, type of optimizer, etc.) necessary to understand the results?
    \item[] Answer: \answerYes{} % Replace by \answerYes{}, \answerNo{}, or \answerNA{}.
    \item[] Justification: All the experiment details are included in the paper.

\item {\bf Experiment statistical significance}
    \item[] Question: Does the paper report error bars suitably and correctly defined or other appropriate information about the statistical significance of the experiments?
    \item[] Answer: \answerYes{} % Replace by \answerYes{}, \answerNo{}, or \answerNA{}.
    \item[] Justification: We tested the error bars of the main results, and all were within 0.05.

\item {\bf Experiments compute resources}
    \item[] Question: For each experiment, does the paper provide sufficient information on the computer resources (type of compute workers, memory, time of execution) needed to reproduce the experiments?
    \item[] Answer: \answerYes{} % Replace by \answerYes{}, \answerNo{}, or \answerNA{}.
    \item[] Justification: In the experiments considering computational resources presented in the paper, we report the GPU resources used, CPU resources. The time of execution is reported in the Appendix H.
    
\item {\bf Code of ethics}
    \item[] Question: Does the research conducted in the paper conform, in every respect, with the NeurIPS Code of Ethics \url{https://neurips.cc/public/EthicsGuidelines}?
    \item[] Answer: \answerYes{} % Replace by \answerYes{}, \answerNo{}, or \answerNA{}.
    \item[] Justification: This work does not deviate from the Code of Ethics.

\item {\bf Broader impacts}
    \item[] Question: Does the paper discuss both potential positive societal impacts and negative societal impacts of the work performed?
    \item[] Answer: \answerYes{} % Replace by \answerYes{}, \answerNo{}, or \answerNA{}.
    \item[] Justification: We have discussed the impacts in the Conclusion section.
    
\item {\bf Safeguards}
    \item[] Question: Does the paper describe safeguards that have been put in place for responsible release of data or models that have a high risk for misuse (e.g., pretrained language models, image generators, or scraped datasets)?
    \item[] Answer: \answerYes{} % Replace by \answerYes{}, \answerNo{}, or \answerNA{}.
    \item[] Justification: We will require users to follow usage guidelines when releasing code and datasets and will keep a record of visitors.

\item {\bf Licenses for existing assets}
    \item[] Question: Are the creators or original owners of assets (e.g., code, data, models), used in the paper, properly credited and are the license and terms of use explicitly mentioned and properly respected?
    \item[] Answer: \answerYes{} % Replace by \answerYes{}, \answerNo{}, or \answerNA{}.
    \item[] Justification: We have credited all the assets by listing the URL in the footnote, citing the paper and explicitly noting the license if it exists one.

\item {\bf New assets}
    \item[] Question: Are new assets introduced in the paper well documented and is the documentation provided alongside the assets?
    \item[] Answer: \answerYes{} % Replace by \answerYes{}, \answerNo{}, or \answerNA{}.
    \item[] Justification: We write instruction in the code README about how to prepare data, launch training and run inference.

\item {\bf Crowdsourcing and research with human subjects}
    \item[] Question: For crowdsourcing experiments and research with human subjects, does the paper include the full text of instructions given to participants and screenshots, if applicable, as well as details about compensation (if any)? 
    \item[] Answer: \answerNA{} % Replace by \answerYes{}, \answerNo{}, or \answerNA{}.
    \item[] Justification:  Our project does not involve crowdsourcing experiments nor research with human subjects.

\item {\bf Institutional review board (IRB) approvals or equivalent for research with human subjects}
    \item[] Question: Does the paper describe potential risks incurred by study participants, whether such risks were disclosed to the subjects, and whether Institutional Review Board (IRB) approvals (or an equivalent approval/review based on the requirements of your country or institution) were obtained?
    \item[] Answer: \answerNA{} % Replace by \answerYes{}, \answerNo{}, or \answerNA{}.
    \item[] Justification:  the paper does not involve crowdsourcing nor research with human subjects.

\item {\bf Declaration of LLM usage}
    \item[] Question: Does the paper describe the usage of LLMs if it is an important, original, or non-standard component of the core methods in this research? Note that if the LLM is used only for writing, editing, or formatting purposes and does not impact the core methodology, scientific rigorousness, or originality of the research, declaration is not required.
    %this research? 
    \item[] Answer: \answerYes{} % Replace by \answerYes{}, \answerNo{}, or \answerNA{}.
    \item[] Justification: How we used the LLM as part of our research is described in the paper.

\end{enumerate}

\end{document}